\documentclass[nofootinbib,preprintnumbers,10pt,twocolumn]{revtex4-1}
\usepackage{amsmath,amssymb,pgf,pgfarrows,pgfnodes,float,appendix, hyperref,slashed,breakurl,graphicx}
\definecolor{Color}{rgb}{0.28, 0.24, 0.55}
\definecolor{Orange}{rgb}{1,0.38,0.11}
\hypersetup{
    colorlinks = true,
    citecolor = Orange,
    linkcolor = Color,
    urlcolor  = Orange,
}
\definecolor{internationalorange}{rgb}{1.0, 0.31, 0.0}

\usepackage{graphicx}
\usepackage{subfigure}
\usepackage[margin=0.9in]{geometry}

\usepackage{comment}

\usepackage{enumerate}

\newcommand{\SU}{\text{SU}}
\newcommand{\SO}{\text{SO}}
\newcommand{\U}{\text{U}}
\newcommand{\vCP}{v_\text{CP}}

\usepackage{color, colortbl}
\definecolor{Gray}{gray}{0.8}
\definecolor{GrayLight}{gray}{0.4}
\definecolor{Darkgreen}{RGB}{30,120,30}
\definecolor{granate}{rgb}{0.8039,0.2,0.2}
\newcommand{\beq}{\begin{equation}}
\newcommand{\eeq}{\end{equation}}
\newcommand{\bea}{\begin{eqnarray}}
\newcommand{\eea}{\end{eqnarray}}
\newcommand{\iu}{\text{i}}
\newcommand{\e}{\text{e}}

\usepackage{tikz}
\usetikzlibrary{"arrows", "automata", "backgrounds", "calendar", "chains", "matrix", "mindmap", "patterns", "petri", "shadows", "shapes.geometric", "shapes.misc", "spy", "trees"}
\usetikzlibrary{arrows.meta, positioning, calc}
\usetikzlibrary{arrows,shapes}
\usetikzlibrary{trees}
\usetikzlibrary{matrix,arrows} 				
\usetikzlibrary{positioning}				
\usetikzlibrary{calc,through}				
\usetikzlibrary{decorations.pathreplacing}  
\usepackage{pgffor}							

\usetikzlibrary{decorations.pathmorphing}	
\usetikzlibrary{decorations.markings}
\tikzset{
    vector/.style={decorate, decoration={snake}, draw},
    scalarloop/.style={dashed,draw=black, postaction={decorate},
        decoration={markings,mark=at position .75 with {\arrow[draw=black]{<}}}},
    scalarloop1/.style={dashed,draw=black, postaction={decorate},
        decoration={markings,mark=at position .25 with {\arrow[draw=black]{<}}}},
	provector/.style={decorate, decoration={snake,amplitude=2.5pt}, draw},
	antivector/.style={decorate, decoration={snake,amplitude=-2.5pt}, draw},
    fermion/.style={draw=black, postaction={decorate},
        decoration={markings,mark=at position .55 with {\arrow[draw=black]{>}}}},
    fermioncyan/.style={draw=black, postaction={decorate},
        decoration={markings,mark=at position .55 with {\arrow[draw=cyan]{<}}}},
    fermiondif/.style={draw=black, postaction={decorate},
        decoration={markings,mark=at position .7 with {\arrow[draw=black]{>}}}},
            fermiondif2/.style={draw=black, postaction={decorate},
        decoration={markings,mark=at position .7 with {\arrow[draw=black]{<}}}},
    fermionend/.style={draw=black, postaction={decorate},
        decoration={markings,mark=at position 1 with {\arrow[draw=black]{>}}}},
    fermionuchannel2/.style={draw=black, postaction={decorate},
        decoration={markings,mark=at position .4 with {\arrow[draw=black]{>}}}},
    scalardif/.style={dashed,draw=black, postaction={decorate},
        decoration={markings,mark=at position .7 with {\arrow[draw=black]{>}}}},
    scalarend/.style={dashed,draw=black, postaction={decorate},
        decoration={markings,mark=at position 1 with {\arrow[draw=black]{>}}}},
    fermionbar/.style={draw=black, postaction={decorate},
        decoration={markings,mark=at position .55 with {\arrow[draw=black]{<}}}},
    fermionnoarrow/.style={draw=black},
    gluon/.style={decorate, draw=black,
        decoration={coil,amplitude=4pt, segment length=5pt}},
    scalar/.style={dashed,draw=black, postaction={decorate},
        decoration={markings,mark=at position .55 with {\arrow[draw=black]{>}}}},
    scalarcyan/.style={dashed,draw=black, postaction={decorate},
        decoration={markings,mark=at position .55 with {\arrow[draw=cyan]{>}}}},
    scalaruchannel1/.style={dashed,draw=black, postaction={decorate},
        decoration={markings,mark=at position .7 with {\arrow[draw=black]{>}}}},
                  scalaruchannel2/.style={dashed,draw=black, postaction={decorate},
        decoration={markings,mark=at position .4 with {\arrow[draw=black]{>}}}},
    scalarbar/.style={dashed,draw=black, postaction={decorate},
        decoration={markings,mark=at position .55 with {\arrow[draw=black]{<}}}},
    scalarnoarrow/.style={dashed,draw=black},
    electron/.style={draw=black, postaction={decorate},
        decoration={markings,mark=at position .55 with {\arrow[draw=black]{>}}}},
	bigvector/.style={decorate, decoration={snake,amplitude=4pt}, draw},
}

\tikzstyle{block} = [draw, rectangle, 
    minimum height=3em, minimum width=6em]

\usepackage{tikz}
\usetikzlibrary{fit}
\tikzset{%
  highlight/.style={rectangle,rounded corners,color=granate,draw,text opacity =1,
    fill opacity=0.5,thick,inner sep=0pt}
}

\usepackage{placeins}

\tikzset{
    cross/.pic = {
    \draw[rotate = 45] (-#1,0) -- (#1,0);
    \draw[rotate = 45] (0,-#1) -- (0, #1);
    }
}

\tikzset{
    square/.style={%
        draw=none,
        circle,
        append after command={%
            \pgfextra \draw[#1] (\tikzlastnode.north-|\tikzlastnode.west) rectangle 
                (\tikzlastnode.south-|\tikzlastnode.east);\endpgfextra}
    },
    square/.default=black
}

\tikzstyle{block} = [draw, rectangle, 
    minimum height=3em, minimum width=6em]

\usepackage{xparse}
\NewDocumentCommand\semiloop{O{black}mmmO{}O{above}}
{%
\draw[#1] let \p1 = ($(#3)-(#2)$) in (#3) arc (#4:({#4+180}):({0.5*veclen(\x1,\y1)})node[midway, #6] {#5};)
}

\begin{document}

\title{A Pati-Salam realization of the Nelson-Barr mechanism}
\author{Clara Murgui}
\affiliation{Theoretical Physics Department, CERN, 1 Esplanade des Particules, CH-1211 Geneva 23, Switzerland}

\preprint{CERN-TH-2026-033}

\begin{abstract}

We present a UV completion of the Standard Model in which quarks and leptons are unified under color $\SU(4)$. A single fermionic representation, the real antisymmetric, provides the building blocks to address the strong CP problem via the Nelson-Barr mechanism, while simultaneously correcting the charged-lepton and down-quark mass relations predicted by Pati-Salam theories for the two heaviest generations. 
We show that the characteristic scales of the theory are strongly constrained by its phenomenology. 
The interplay between the quality of the Nelson-Barr mechanism and the non-observation of baryon-number-violating processes determines the scale of spontaneous CP violation and the mass of the new vector-like down quark. The theory predicts a distinctive baryon-number-violating decay mode of the neutron, $n \to K^+ \ell^-$ (with $\ell = e,\mu$),
 which lies within the projected sensitivity of upcoming nucleon-decay experiments such as Hyper-Kamiokande, JUNO, and DUNE.  
\end{abstract}

\maketitle

\section{Introduction}

The fact that, in the presence of an ${\cal O}(1)$ charge-parity (CP)-violating phase in the flavor sector~\cite{ParticleDataGroup:2024cfk}, the combination inducing a nonzero electric dipole moment (EDM) in hadrons 
is strongly constrained by null measurements of the neutron EDM~\cite{Abel:2020pzs,Pospelov:1999mv,Liang:2023jfj}, 
\begin{equation}
\bar{\theta}_{\rm QCD} = \theta_{\rm QCD} + \arg \{ \det ({\cal M}_u {\cal M}_d)\} < 10^{-10}~,
\label{eq:SCPproblem}
\end{equation} 
constitutes a long-standing puzzle in the Standard Model (SM) of particle physics.

A popular resolution is to dynamically set $\bar \theta_{\rm QCD}$ to zero by introducing a global symmetry that is anomalous under QCD~\cite{Peccei:1977hh,Peccei:1977ur}. The associated pseudo-Goldstone boson, the axion~\cite{Wilczek:1977pj,Weinberg:1977ma}, relaxes $\bar \theta_{\rm QCD}$ to its CP-conserving minimum in the infrared, while potentially accounts for the observed dark matter abundance in the universe~\cite{Preskill:1982cy,Abbott:1982af,Dine:1982ah}. However, the Peccei-Quinn mechanism generically suffers from severe quality problems~\cite{Georgi:1981pu}: global symmetries are not expected to be respected in the ultraviolet (UV), and higher-dimensional operators can shift the axion potential minimum, inducing an effective $\bar \theta_{\rm QCD}$ that violates the bound in Eq.~\eqref{eq:SCPproblem}. Various approaches have been proposed to address this issue, including heavy axion models ({\it e.g.}~\cite{Gherghetta:2016fhp,Gaillard:2018xgk}), constructions enforcing high-quality global symmetries ({\it e.g.}~\cite{Duerr:2017amf,Agrawal:2025mke}), and mechanisms arising in higher-dimensional or string-inspired axiverse scenarios ({\it e.g.}~\cite{Choi:2003wr,Svrcek:2006yi,Arvanitaki:2009fg}).

There are, however, alternative ways to understand the smallness of $\bar \theta_{\rm QCD}$. From Eq.~\eqref{eq:SCPproblem}, it follows that in a CP-invariant theory the topological $G\tilde G$ term is forbidden in the Lagrangian and the Yukawa couplings are real, implying $\bar \theta_{\rm QCD}=0$. However, CP violation is known to be present in the Standard Model, as evidenced by the non-vanishing Jarlskog invariant~\cite{ParticleDataGroup:2024cfk}. To account for that, one can assume that CP is a symmetry of the Lagrangian but it is spontaneously broken by the vacuum of the theory. In this context, the quark mass matrices can be complex and, as long as they satisfy
\begin{equation}\label{eq:argdet0}
    \arg\{\det({\cal M}_u {\cal M}_d) \} = 0~,
\end{equation}
the strong CP problem can be resolved consistently with the presence of an ${\cal O}(1)$ CP-violating phase in the flavor sector.

The above condition can be naturally realized through suitable textures of the quark mass matrices.\footnote{In addition to the Nelson-Barr mechanism, on which we focus in this paper, there exist other mechanisms leading to Eq.~\eqref{eq:argdet0}; see, for example, Refs.~\cite{Mohapatra:1978fy,PhysRevD.41.1286}.} Nelson and Barr~\cite{Nelson:1983zb,Barr:1984qx,Nelson:1984hg} proposed extending the SM by vector-like quarks that couple only to the right-handed colored sector through a scalar whose vacuum expectation value (vev) breaks CP. In such constructions, quark mass matrices of the form 
\begin{equation}\label{eq:texture}
    {\cal M}_q = \begin{pmatrix} \mathsf{M}_q & 0 \\ \mu_i & M_Q \end{pmatrix}~,
\end{equation}
where $\mathsf{M}_q$ denotes the $3\times 3$ SM weak mass matrix (real), $M_Q$ the mass of the new vector-like quark (real), and $\mu_i$ the mixing mass terms between the SM quarks and the new state (complex), 
allow for ${\cal O}(1)$ CP violation in the Cabibbo-Kobayashi-Maskawa (CKM) matrix while maintaining $\bar \theta_{\rm QCD} = 0$ at tree level. 

The minimal simplified model realizing this idea was proposed by Bento, Branco and Parada (BBP) in Ref.~\cite{Bento:1991ez}. In this construction, a vector-like down-type quark, $D_{L,R} \sim(3,1,-1/3)_{\rm SM}$, is introduced together with a complex scalar $S \sim (1,1,0)_{\rm SM}$, both charged under a $\mathbb{Z}_2$ symmetry. The subscript ``SM" here and in the following denotes the quantum numbers under the SM gauge group ${\cal G}_{\rm SM} = \SU(3)_C \otimes \SU(2)_L \otimes \mathrm{U}(1)_Y$. The relevant Lagrangian in this case is given by,
\begin{equation}
\begin{split}
   - {\cal L}& \supset \mathsf{Y}_d \, \bar Q_L H d_R + M_D \, \bar D_L D_R \\
   & \quad  + \bar D_L (\lambda \, S + \lambda' S^*) \, d_R+   \text{h.c.}~,
\end{split}
\end{equation}
where $Q_L \sim(3,2,1/6)_{\rm SM}$, $d_R\sim(3,1,-1/3)_{\rm SM}$, and $H\sim(1,2,1/2)_{\rm SM}$ are SM fields, all singlets under the new $\mathbb{Z}_2$. In the broken phase, $\mu_i = \lambda_i \langle S \rangle \in \mathbb{C}$, and the texture in Eq.~\eqref{eq:texture} is realized, leading to the prediction $\bar \theta_{\rm QCD} = 0$ at tree level.

However, higher dimensional operators, appearing already at dimension five, 
\begin{equation}\label{eq:dangerousO}
    \frac{1}{\Lambda}\bar Q_L H  (S D_R) + \cdots~,
\end{equation}
reintroduce a small $\bar \theta_{\rm QCD}$ and spoil the mechanism unless the scale of spontaneous CP violation is below $40$ TeV~\cite{Murgui:2025scx}, assuming $\Lambda \sim M_{\rm Pl} = 1.2 \times 10^{19}$ GeV. Such a low scale is incompatible with high-scale leptogenesis~\cite{Asadi:2022vys}, which requires the relevant lepton-number- and CP-violating dynamics to occur at
a much higher scale, and with inflationary models featuring a hot reheating phase, since the domain walls arising from the spontaneous breaking of CP must be inflated away~\cite{McNamara:2022lrw}. 

Automatic Nelson-Barr models have been proposed~\cite{Asadi:2022vys,Perez:2023zin,Murgui:2025scx}, in which the $\mathbb{Z}_2$ is an accidental remnant of the gauge symmetry and matter content. This ensures that the operator in Eq.~\eqref{eq:dangerousO} is generically forbidden and contributions to $\bar \theta_{\rm QCD}$ are further suppressed, thereby allowing for a higher spontaneous CP violation scale~\cite{Asadi:2022vys}.

In this paper, we show that the Nelson-Barr mechanism can be automatically realized in simple and well-motivated extensions of the SM. In particular, we provide a UV completion of the minimal BBP simplified model~\cite{Bento:1991ez} in the context of quark-lepton unification~\cite{Pati:1974yy}. The framework treats quarks and leptons on equal footing, explains charge quantization, naturally includes right-handed neutrinos, and leads to approximate mass relations between the down-quarks and the charged leptons.
 
Within color $\SU(4)$, the next-to-minimal nontrivial representation beyond the fundamental that can be added to the matter content is the antisymmetric representation, $F_6$. Remarkably, this representation is real and therefore does not spoil anomaly cancellation. It contains a vector-like down quark, $D_{L,R}$, which will play a two-fold role: (i) together with the $\SU(4)_C$ symmetry and the scalars responsible for spontaneous CP breaking, it leads to the automatic generation of the mass texture required for the Nelson-Barr mechanism shown in Eq.~\eqref{eq:texture}, and (ii) its mixing with the SM down-type quarks helps correct the Pati-Salam mass matrix relation $\mathsf{M}_e= \mathsf{M}_d$ for the second and third generations. As we will show, in this framework the scale of quark-lepton unification is bounded by an exotic baryon-number-violating decay channel of the neutron, for which future experimental sensitivities provide a test of the theory.

The paper is organized as follows. We first introduce the model in Sec.~\ref{sec:TF}. In Sec.~\ref{sec:strongCP} we show how it addresses the strong CP problem, while in Sec.~\ref{sec:masses}, we discuss the origin of fermion masses and of an ${\cal O}(1)$ CP-violating phase. Section~\ref{sec:BNV} is devoted to the various sources of baryon-number violation and their phenomenological implications in relation to the other relevant scales of the theory. In Sec.~\ref{sec:Scales}, we combine the results derived throughout the paper and discuss the resulting hierarchy of scales.
The first two figures in this manuscript provide a compact summary of the dynamics of the theory: Fig.~\ref{fig:plot} shows the allowed parameter space for the new vector-like quark mass and the scale of spontaneous CP violation, while Fig.~\ref{fig:summary} illustrates the hierarchy of scales in this simple UV completion of the SM. Fig.~\ref{fig:GUT} shows the possibility of gauge-coupling unification in the theory. We conclude in Sec.~\ref{sec:remarks}.

\section{Quark-Lepton unification}
\label{sec:TF}
A simple extension of the SM gauge symmetry that realizes quark-lepton unification is given by~\cite{Smirnov:1995jq,FileviezPerez:2013zmv}
\begin{equation}
{\cal G}_{\rm PS} = \SU(4)_C \otimes \SU(2)_L \otimes \mathrm{U}(1)_R,
\end{equation}
which is anomaly free for the matter content:
\begin{equation}\label{eq:matter}
\begin{split}
&F_{QL} \sim (4,2,0)_{\rm PS} = (Q_L^\alpha, L_L)^T~,\\ 
&F_{u\nu} \sim (4,1,1/2)_{\rm PS} = (u_R^\alpha, \nu_R)^T~,\\
&F_{de}  \sim (4,1,-1/2)_{\rm PS} = (d_R^\alpha, e_R)^T~.
\end{split}
\end{equation}
Here, $u_R \sim (3,1,2/3)_{\rm SM}$, $e_R \sim (1,1,-1)_{\rm SM}$, $L_L\sim (1,2,-1/2)_{\rm SM}$, and $\nu_R \sim (1,1,0)_{\rm SM}$, while $Q_L$ and $d_R$ have been defined previously. The index $\alpha=1,2,3$ denotes color.
In Eq.~\eqref{eq:matter} and in the following, the subscript ``PS'' refers to the quantum numbers under ${\cal G}_{\rm PS}$. The three representations above constitute one SM family plus a right-handed neutrino. Therefore, three copies of them, together with a scalar boson $H \sim (1,2,1/2)_{\rm PS}$, are needed to accommodate the full SM field content.

We assume that ${\cal G}_{\rm PS}$ breaks to the SM in two steps, triggered by the vevs of the adjoint $\Phi_{15} \sim (15,1,0)_{\rm PS}$ and symmetric $\Phi_{10}\sim (10,1,1)_{\rm PS}$ scalar representations, which decompose under ${\cal G}_{\rm SM}$ as 

\begin{equation}\label{eq:scalars}
\begin{split}
 \!\! \Phi_{15} & \!=\! \begin{pmatrix} \phi_{15}^{\alpha \beta} & \! \!\! \phi_{15}^\alpha \\\! (\phi_{15}^\alpha)^* &\!\!\!  0  \end{pmatrix}\!  + T_{15} \phi_{15}^0 , \, \Phi_{10} \!=\! \begin{pmatrix} \phi_{10}^{\alpha \beta} & \frac{\phi_{10}^\alpha}{\sqrt{2}} \\  \frac{\phi_{10}^\alpha}{\sqrt{2}} & \phi_{10}^0 \end{pmatrix},
\end{split}
\end{equation}
with $\phi_{15}^0 \sim (1,1,0)_{\rm SM}$ and $\phi_{10}^0 \sim (1,1,0)_{\rm SM}$ being singlets under the SM gauge group. 
The breaking ${\cal G}_{\rm PS} \to {\cal G}_{\rm SM}$ proceeds according to the chain
\begin{equation}
\begin{split}
    {\cal G}_{\rm PS} &\stackrel{\langle \Phi_{15}\rangle}{\to} \SU(3)_C \otimes \mathrm{U}(1)_{B-L} \otimes \SU(2)_L   \otimes \mathrm{U}(1)_R\\
    &\stackrel{\langle \Phi_{10}\rangle}{\to} {\cal G}_{\rm SM}.
    \end{split}
\end{equation}
In particular, the vev of the adjoint representation, $\langle \Phi_{15} \rangle = v_{15}\,T_{15}$, breaks $\SU(4)_C \to \SU(3)_C \otimes \mathrm{U}(1)_{B-L}$, where the unbroken abelian factor is generated by the $\SU(4)_C$ Cartan generator
\begin{equation}
T_{15} = \frac{1}{2\sqrt{6}}\text{diag}(1,1,1,-3),
\end{equation}
corresponding to the charge assignment
\begin{equation}
{\cal Q}_{B-L} = \frac{2\sqrt{6}}{3}T_{15}~.
\end{equation}

 The vev of the symmetric representation, $\langle \Phi_{10}^{AB} \rangle = \delta^{A4}\delta^{4B} (v_{10}/\sqrt{2})$, then spontaneously breaks a linear combination of $\mathrm{U}(1)_{B-L} \otimes \mathrm{U}(1)_R$ to the SM hypercharge, given by
\begin{equation}\label{eq:Ynorm}
    {\cal Q}_Y = {\cal Q}_R + {\cal Q}_{B-L}/2~.
\end{equation}

Apart from the two breaking scales $v_{15}$ and $v_{10}$, the gauge sector is fully determined by matching to the SM inputs at the relevant scale.
The vector leptoquarks associated with the broken generators of $\SU(4)_C$, $X_\mu \sim (3,1,2/3)_{\rm SM}$, and the neutral gauge boson associated with the combination orthogonal to the SM hypercharge,  $Z'\sim(1,1,0)_{\rm SM}$, acquire masses 
\begin{equation}
M_X^2 \! = \! \frac{g_4^2}{2} \left( \!\frac{4}{3}v_{15}^2+v_{10}^2\!\right)\!,\ M_{Z'}^2 \! = \! \left(\! g_R^2 + \frac{3}{2}g_4^2\! \right) v_{10}^2.
\end{equation}
At the quark-lepton unification scale, the gauge couplings are given by
\begin{equation}
    g_4 = \sqrt{4\pi \alpha_s}~, \quad g_R = \frac{\sqrt{4\pi \alpha_{\rm em}}}{\cos \theta_W \cos \theta_{QL}}~,
\end{equation}
while the $\SU(2)_L$ gauge coupling is $g_2 =   \sqrt{4\pi \alpha_{\rm em}}/\sin \theta_W$, with $\theta_W$ being the Weinberg angle and $\alpha_{\rm em}$ the fine structure constant. 
The quark-lepton unification angle $\theta_{\rm QL}$, 
\begin{equation}\label{eq:thetaQL}
    \sin^2\theta_{QL} = \frac{g_R^2}{g_R^2+ \frac{3}{2} g_4^2} =  \frac{2}{3}\frac{\alpha_{\rm em}}{\alpha_s} \frac{1}{\cos^2 \theta_W}~,
\end{equation}
parametrizes the mixing between the $\mathrm{U}(1)_R$ and $\mathrm{U}(1)_{B-L}$ generators. Using these relations, the gauge boson masses can be written as 
\begin{equation}
\begin{split}
M_X^2 &= 2\pi \alpha_s\bigg(\frac{4}{3}v_{15}^2 + v_{10}^2\bigg),   \\
M_{Z'}^2 &= 6 \pi \alpha_s \bigg(\frac{1}{1-\sin^2 \theta_{\rm QL}}\bigg)v_{10}^2~.
\end{split}
\end{equation}
 For later reference, we identify the quark-lepton unification scale as $\Lambda_{\rm QL}  \sim M_X \sim v_{15}$.

Electroweak symmetry breaking occurs in the usual manner through the vev of the SM Higgs boson, $H\sim (1,2,1/2)_{\rm PS}$. However, owing to the reduced number of matter representations of the theory, only two Yukawa couplings are allowed, 
\begin{equation}\label{eq:Yukawa}
-{\cal L} \supset \mathsf{Y}_{u\nu} \bar F_{QL} i \sigma_2 H^* F_{u\nu} + \mathsf{Y}_{de} \bar F_{QL} H F_{de} + \text{h.c.}~,
\end{equation}
which lead to the following relations among the fermion mass matrices:
\begin{equation}
    \mathsf{M}_u = \mathsf{M}_\nu^D, \qquad \text{and} \qquad \mathsf{M}_d = \mathsf{M}_e.
\end{equation}
The relation between the up-type quark and neutrino mass matrices implies that neutrinos must be Majorana particles; otherwise, the resulting neutrino masses would be phenomenologically unacceptable. For the symmetry-breaking pattern considered here, right-handed neutrinos acquire a Majorana mass proportional to the $\mathrm{U}(1)_{B-L}\otimes \mathrm{U}(1)_R \to \mathrm{U}(1)_Y$ breaking scale through the renormalizable interaction
\begin{equation}\label{eq:Majorana}
-{\cal L} \supset \mathsf{Y}_{\nu R} \, \Phi_{10}^* F_{u\nu}^T C F_{u \nu} + \text{h.c.}~,
\end{equation}
where $C$ denotes the charge-conjugation matrix.

We note that, in this Type-I seesaw~\cite{Minkowski:1977sc,Gell-Mann:1979vob,Mohapatra:1979ia,Yanagida:1979as} the Dirac mass matrix, $\mathsf{M}_\nu^D$, which plays the role of the {\it pivot} of the seesaw, is fixed by the up-type quark masses. As a result, the quark-lepton unification scale must be sufficiently large in order to generate active neutrino masses below the experimental upper bound $m_\nu  < 0.1$ eV~\cite{Planck:2018vyg,KATRIN:2021uub,DESI:2024mwx}. This, in turn, places a lower bound on the quark-lepton unification scale at the canonical seesaw scale, $\Lambda_{\rm QL} \gtrsim v_{10} \sim 10^{14}$ GeV.

The mass matrix relation $\mathsf{M}_d=\mathsf{M}_e$, while working approximately for the third generation after renormalization-group running~\cite{Georgi:1979df}, fails to reproduce the masses of the lighter families. 
Given the high scale of $\langle \phi_{15}^0 \rangle \gtrsim 10^{14}$ GeV, one must therefore consider the effect of the lowest-dimensional non-renormalizable operators such as
\begin{equation}\label{eq:nonrensplitting}
-{\cal L}\supset \frac{1}{M_{\rm Pl}} \bar F_{\rm QL} H \Phi_{15} F_{de} + \text{h.c.}~.
\end{equation}
Since $\langle \Phi_{15} \rangle \propto v_{15} \, \text{diag}(1,1,1,-3)$, this operator splits the down-type quark and charged lepton masses, effectively acting as a second Higgs doublet\footnote{In conventional Pati-Salam constructions, a scalar multiplet in the fundamental of $\SU(2)_L$ and in the adjoint of $\SU(4)_C$ is often introduced to generate the desired mass splitting.}.
The resulting correction is of order $v_{15}/M_{\rm Pl} \sim {\cal O}(\text{MeV})$, which is naturally of the order of the down-quark mass. However, this contribution alone is insufficient to account for realistic fermion masses for the second and third generations, indicating that a modest extension of the theory is still required.

In this work, we extend the matter content of the quark-lepton unification theory by introducing the simplest nontrivial real fermionic representation of $\SU(4)_C$, 
\begin{equation}
\begin{split}
F_6 &= \frac{1}{\sqrt{2}}\begin{pmatrix} \epsilon^{\alpha \beta \gamma} (D^c)_{L\gamma} & D_L^\alpha \\ - D_L^\alpha & 0 \end{pmatrix} \sim (6,1,0)_{\rm PS}~,
\end{split}
\end{equation}
which contains a vector-like down quark, $D_L \sim (3,1,-1/3)_{\rm SM}$ and $(D^c)_L = (D_R)^c \sim (\bar 3,1,1/3)_{\rm SM}$. In addition, we introduce two scalar fields in the fundamental representation of $\SU(4)_C$, $\Phi_4 = (\phi_4^\alpha,\phi_4) \sim (4,1,1/2)_{\rm PS}$. Within this matter content, Lorentz invariance and the ${\cal G}_{\rm PS}$ gauge symmetry allow the following interactions at the renormalizable level:
\begin{equation}\label{eq:F6Yukawa}
\begin{split}
- {\cal L} &\supset \tfrac{1}{2}M_6 (F_6^{AB})^T C  \, F_6^{CD} \epsilon_{ABCD} \\
&\quad +  \bar F_{de} F_6 (\lambda_a \Phi_{4a}^* + \lambda_b \Phi_{4b}^*)+ \text{h.c.}~,
\end{split}
\end{equation}
 which, on the one hand, enable realistic mass relations for the heavier fermion generations through the mixing of the vector-like quark with the SM down-type quarks and, on the other hand, provide a realization of the Nelson-Barr mechanism for solving the strong CP problem, provided CP is an exact symmetry of the ultraviolet Lagrangian.

\section{The strong CP problem}
\label{sec:strongCP}
Let us assume that CP is an exact symmetry of the Lagrangian, so that $\theta_{\rm QCD} = 0$ in Eq.~\eqref{eq:SCPproblem}. Although all parameters in the Lagrangian  are real, the vacuum of the theory is not invariant under CP due to a nonzero phase generated by a linear combination of the $\Phi_4$ fields in the scalar potential.\footnote{The CP phases potentially induced by the vacuum expectation values of the scalar fields $\Phi_{15}$, $\Phi_{10}$, and 
$H$ are unphysical, as they can be removed by gauge transformations associated with the generators broken by the corresponding vevs.} This is illustrated by the following terms in the scalar potential:
\begin{equation}\label{eq:potentialCPphase}
V \supset   -\tilde \mu_4^2 \Phi_{4a} \Phi_{4b}^* + \tilde \lambda_4 (\Phi_{4a}^*\Phi_{4b})^2+ \text{h.c.}~.
\end{equation}
After the $\Phi_{4}$ fields acquire vevs, $\langle \Phi_{4a}^A \rangle= e^{i \theta_{a}} v_{4_{a}} \delta^{A4}/ \sqrt{2}$ (and similarly for $\langle \Phi_{4b} \rangle$), the above terms can be rewritten as 
\begin{equation}
V\supset - \tilde \mu^2_4 v_{4a} v_{4b} \cos \theta_4 + \tilde \lambda_4 \frac{v_{4a}^2 v_{4b}^2}{2} \cos 2 \theta_4~,
\end{equation}
with $\theta_4=\theta_{4a} - \theta_{4b}$. Minimizing the potential with respect to $\theta_4$ fixes the surviving CP-violating phase as
\begin{equation}
\cos \theta_4 =  \frac{\tilde \mu_4^2}{2 \tilde \lambda_4 v_{4a} v_{4b} }~.
\label{eq:CPphase}
\end{equation}
We note that the accidental $\mathrm{U}(1)_{\Phi_{4a} - \Phi_{4b}}$ symmetry in Eq.~\eqref{eq:potentialCPphase} is explicitly broken by additional terms in the scalar potential, such as
\begin{equation}\label{eq:mix410}
    V \supset f_1 \Phi_{4a}^2 \Phi_{10}^* + f_2 \Phi_{4a}\Phi_{4b} \Phi_{10}^*  + \cdots~,
\end{equation}
which generically modify the relation in Eq.~\eqref{eq:CPphase}.

The physical phase $\theta_4$ renders the $4\times 4$ down-type quark mass matrix complex, 
\begin{equation}\label{eq:Md}
{\cal M}_d = \begin{pmatrix} \displaystyle (\mathsf{Y}_{de})^{ij} \frac{v_H}{\sqrt{2}} & & 0^{i4}  \\ \displaystyle \tfrac{1}{2}(\lambda_a^j v_{4a}  \e^{\iu \theta_{4a}} + \lambda_b^j v_{4b} \e^{\iu \theta_{4b}} )&  & M_6 \end{pmatrix}~,
\end{equation}
where $\langle H^i \rangle = \delta^{i2} v_H/\sqrt{2}$ is real, with $v_H = 246$ GeV. Roman indices $i,j=1,2,3$ run over the three SM families. The matrix ${\cal M}_d$ is written in the flavor basis $(d,s,b,D)$. For later convenience, we parametrize the linear combination of the entries ${\cal M}_d^{4j}\equiv \lambda_j \vCP \, \e^{\iu \theta_j}/2$. 
The structure of Eq.~\eqref{eq:Md} allows the CKM matrix to contain a physical CP-violating phase while ensuring $\arg \{ \det ({\cal M}_d) \} = 0$. The latter property can be seen explicitly by expanding the determinant along the fourth column. As a result, $\bar \theta_{\rm QCD} = 0$ at tree level. 

Additional contributions to $\bar \theta_{\rm QCD}$ arise nevertheless from radiative corrections and higher dimensional operators involving $\theta_4$ that enter the down-type quark masses~\cite{Dine:2015jga}. 

At one loop, processes such as the one illustrated below generate complex contributions to the down-type quark mass matrix in the broken phase~\cite{Asadi:2022vys,Perez:2023zin},
\begin{equation*}
\begin{gathered}
\begin{tikzpicture}[line width=1.5 pt,node distance=1 cm and 1 cm]
\coordinate[label=left:$b_R$](dR);
\coordinate[right = 1 cm of dR](v1);
\coordinate[right = 2 cm of v1](v2);
\coordinate[right= 1 cm of v2,label=right:$b_L$](dL);
\coordinate[right= 1 cm of v1](vaux);
\coordinate[below =  0.75 cm of vaux,label=below:$\langle \Phi_{4a,b} \rangle$](Xk);
\coordinate[above = 1 cm of vaux](v4);
\coordinate[above left = 1 cm of v4,label=left:$\langle \Phi_{4a} \rangle$](Xi);
\coordinate[above right = 1 cm of v4,label=right:$\langle H \rangle$](Xj);
\coordinate[right = 0.5cm of v1,label=below:$D_L$](DLlabel);
\coordinate[right = 0.5cm of vaux,label=below:$b_R$](dRlabel);
\coordinate[above = 0.75 cm of v1,label=left:$\Phi_{4b}$](Xlabel);
\coordinate[above = 0.75 cm of v2,label=right:$H$](Xlabel);
\draw[scalarnoarrow] (v4)--(Xi);
\draw[scalarnoarrow] (v4)--(Xj);
\draw[fermion](dR)--(v1);
\draw[fermion](v1)--(vaux);
\draw[fermion](vaux)--(v2);
\draw[fermion](v2)--(dL);
\draw[scalarnoarrow](vaux)--(Xk);
\draw[fill=black] (v1) circle (.05cm);
\draw[fill=black] (v2) circle (.05cm);
\draw[fill=black] (vaux) circle (.05cm);
\draw[fill=black] (v4) circle (.05cm);
\semiloop[scalarloop]{v1}{v2}{0};
\semiloop[scalarloop1]{v1}{v2}{0};
\end{tikzpicture}
\end{gathered}~,
\end{equation*}
which are suppressed by the quartic interaction $\lambda_{H\Phi} H^\dagger H \Phi_{4a}^\dagger \Phi_{4b}$ and by a loop factor. The resulting shift in $\bar \theta_{\rm QCD}$ can be estimated as
\begin{equation}\label{eq:radcon}
    \Delta \bar \theta_{\rm QCD} = \frac{\lambda_{H\Phi}}{32\pi^2} \frac{\lambda_3^2\vCP^2}{M_{\Phi_4}^2}  \ln \left(\frac{M_{\Phi_4}^2}{m_b^2} \right)~,
\end{equation}
where $\lambda_3$ denotes the mixing coupling involving the bottom quark. 
We note that even when $\lambda_3 ={\cal O}(1)$, the quartic coupling alone can ensure a sufficiently small radiative contribution provided $\lambda_{H\Phi} < 10^{-9}$. This coupling can itself be radiatively generated at one loop by the mixing between the new vector-like quark and the SM down quarks, 
\begin{equation}
\lambda_{H\Phi}|_{{\rm 1-loop}}\sim \bigg(\frac{y_b}{4\pi}\lambda_3 \bigg)^2~,
\end{equation}
where $y_b$ is the bottom-quark yukawa coupling.
 Even in this case, Eq.~\eqref{eq:radcon} is naturally consistent with the experimental bound $\bar \theta_{\rm QCD} < 10^{-10}$, given that the couplings  $\lambda_j$ remain perturbative.

We close this section by commenting on potential shifts on $\bar \theta_{\rm QCD}$ induced by higher-dimensional operators. Because the $F_6$ representation is real, it admits a mass term in the Lagrangian through a super-renormalizable interaction. Consequently, the leading corrections to $\bar \theta_{\rm QCD}$ from higher-dimensional operators arise already at dimension five, through operators of the form
\begin{equation}\label{eq:dim5op}
    \frac{1}{\Lambda_{\rm PS}} (F_6^{AB})^TC\,F_6^{CD} \epsilon_{ABCD} \, \Phi_{4a}\Phi_{4b}^*.
\end{equation}
Requiring the induced shift in $\bar \theta_{\rm QCD}$ to remain acceptably small,
\begin{equation}
    \Delta \bar \theta_{\rm QCD}  \simeq \text{Im}\{\text{Tr}\{ {\cal M}_d^{-1} \delta {\cal M}_d \}\} < 10^{-10},
\end{equation}
one obtains the following upper bound on the spontaneous CP breaking scale~\cite{Asadi:2022vys,Murgui:2025scx}
\begin{equation}\label{eq:boundquality}
    \vCP <  2\times 10^{6} \, \text{GeV} \left(\frac{\Lambda_{\rm PS}}{M_{\rm Pl}}\right)^{\tfrac{1}{2}} \!\! \left(\frac{M_6}{\text{TeV}}\right)^{\tfrac{1}{2}}\!\!\left(\frac{\Delta \bar \theta_{\rm QCD}}{10^{-10}}\right)^{\tfrac{1}{2}}\!\! .
\end{equation}
As pointed out in Ref.~\cite{Asadi:2022vys}, domain walls generated by the breaking of a gauged discrete symmetry require a mechanism to inflate them away. In order for inflation to successfully remove these defects without subsequently restoring CP, the reheating temperature must satisfy $T_{\rm RH} \lesssim v_{\rm CP}$. Many simple realizations of inflation feature large reheating temperatures, which in turn favors larger values of $v_{\rm CP}$. As we will show in Sec.~\ref{sec:BNV}, a large $v_{\rm CP}$ is also required by bounds on baryon number violating nucleon decays in this model. Ultimately, for a vector-like down-quark mass of order $\vCP$, Eq.~\eqref{eq:boundquality} gives the following upper bound: 
\begin{equation}\label{eq:v4bound}
M_6 \sim v_{\rm CP} <   10^{10}\,\text{GeV}.
\end{equation}

In the next section we discuss the impact of the mixing in Eq.~\eqref{eq:Md} on the fermion mass spectrum and explain how an ${\cal O}(1)$ CKM phase is generated.

\section{Fermion masses and mixing}
\label{sec:masses}
As outlined in Sec.~\ref{sec:TF}, the prediction for the mass-matrix relation $\textsf{M}_u = \textsf{M}_\nu^D$ involves the Dirac masses of the neutrinos. The vev of $\Phi_{10}$, responsible for the breaking of $\mathrm{U}(1)_{B-L} \otimes \mathrm{U}(1)_R \to \mathrm{U}(1)_Y$, generates Majorana masses for the right-handed neutrinos, $\mathsf{M}_{\nu_R} = \mathsf{Y}_{\nu R} \, v_{10} / \sqrt{2}$, via the interaction in Eq.~\eqref{eq:Majorana}. Active neutrinos then acquire masses through the type-I seesaw mechanism, $m_\nu \sim m_t^2/v_{10}$.
Because the Dirac neutrino masses are fixed by the up-type quark masses, accommodating neutrino oscillation data~\cite{Esteban:2018azc} as well as cosmological~\cite{Planck:2018vyg,DESI:2024mwx} and laboratory~\cite{KATRIN:2021uub} constraints requires a high scale for $M_{\nu_R}$, corresponding to $v_{10}\sim 10^{14}$ GeV.

The mass-matrix relation $\textsf{M}_d = \textsf{M}_e$ applies to the three SM  generations at the quark-lepton unification scale. However, as is evident from Eq.~\eqref{eq:Md}, the presence of a new vector-like down quark modifies this relation through its mixing with the SM down-type quarks. Diagonalizing the hermitian matrix ${\cal M}_d {\cal M}_d^\dagger$, with ${\cal M}_d$ defined in Eq.~\eqref{eq:Md}, provides both the physical masses of the down-type quarks via the eigenvalues and the CP-violating phase of the CKM matrix via the eigenvectors.

Let us evaluate the down-type quark mass matrix ${\cal M}_d$ at the quark-lepton unification scale, $\Lambda_{\rm QL}$. Following the procedure introduced in Ref.~\cite{Perez:2023zin}, we rotate the SM down-type quarks to the basis in which the $3\times 3$ SM Yukawa block is diagonal. In this basis, the full $4\times 4$ mass matrix takes the form
\begin{equation}\label{eq:MdO}
O_{L}^T {\cal M}_d O_{R} = \begin{pmatrix} m_e  & 0 & 0 & 0 \\ 0 &  m_\mu  & 0 & 0 \\ 0 & 0 &  m_\tau  & 0 \\ \mu_1 & \mu_2 & \mu_3 & M_6 \end{pmatrix},
\end{equation}
where the orthogonal matrices $O_{L,R}$ act trivially on the vector-like quark entry, i.e. $O_{L,R}^{A4} = O_{L,R}^{4A} = 0$, and $O_{L,R}^{44}=1$. The mixing parameters are given by
\begin{equation}
    \mu_j \equiv \frac{1}{2} \sum_{k=1}^3 \left( \lambda_a^k v_{4a} \e^{\iu \theta_{4a}} + \lambda_b^k v_{4b} \e^{\iu \theta_{4b}} \right) \mathsf{O}_{R}^{kj}~,
\end{equation}
with $\textsf{O}_{L,R}$ being the $3\times 3$ block $O_{L,R}^{ij}$. In writing Eq.~\eqref{eq:MdO}, we have used the quark-lepton unification relation $\textsf{M}_d (\Lambda_{\rm QL}) = \textsf{M}_e(\Lambda_{\rm QL})$.

Although the masses appearing in the upper $3\times 3$ block of Eq.~\eqref{eq:MdO} are not yet the physical masses of the SM down-type quarks, they do correspond to the physical charged-lepton masses, since $\textsf{O}_L^T \textsf{M}_e \textsf{O}_R = \text{diag}(m_e, m_\mu, m_\tau)$. 

Below the Pati–Salam breaking scale, the unified Yukawa coupling 
$\mathsf{Y}_{de}$	
  splits into the Standard Model Yukawa matrices for down-type quarks and charged leptons. We denote these by $\mathsf{Y}_d$ and $\mathsf{Y}_e$, respectively, which satisfy the matching condition $\mathsf{Y}_d(\Lambda_{\rm QL})= \mathsf{Y}_e(\Lambda_{\rm QL})=\mathsf{Y}_{de}(\Lambda_{\rm QL})$. To estimate how these Yukawa matrices evolve toward low energies, we consider the dominant contribution to the renormalization-group running of the down-type Yukawa couplings,
\begin{equation}
\begin{split}
 \frac{\text{d} \mathsf{Y}_d(\mu)}{\text{d}\ln \mu} &\simeq -\frac{2}{\pi}\alpha_s(\mu)\mathsf{Y}_d(\mu)~,
\end{split}
\end{equation}
which arises from gluon quantum corrections, while in this approximation $\text{d}Y_e / \text{d}\ln \mu = 0$.  On the other hand, the running of the strong coupling is governed by
\begin{equation}
\frac{\text{d}\alpha_s(\mu)}{\text{d}\ln \mu} \simeq -\bigg( 11- \frac{2}{3}n_f \bigg)\frac{\alpha_s^2(\mu)}{2\pi}~,
\end{equation}
with $n_f$ denoting the number of active quark flavors. Integrating these equations yields the approximate relation,
\begin{equation}
\frac{\mathsf{M}_d^{ij}(\mu)}{\mathsf{M}_d^{ij}(\Lambda_{\rm QL})} = \bigg(\frac{\alpha_s(\mu)}{\alpha_s(\Lambda_{\rm QL})} \bigg)^{\frac{4}{11-2n_f/3}}~.
\end{equation}
Taking threshold effects into account, the down-type quark masses at low energies satisfy $\mathsf{M}_d (2 \text{ GeV}) = \kappa_M \, \mathsf{M}_d(\Lambda_{\rm QL}) = \kappa_M \, \mathsf{M}_e (\Lambda_{\rm QL} )$,  with $\kappa_M \simeq 3.6$  when only one-loop QCD effects are included. Once the contributions proportional to the top Yukawa coupling are taken into account, which enter with the opposite sign relative to the gauge contribution in the renormalization-group equations, this factor is reduced to $\kappa_M \simeq 3$~\cite{Georgi:1979df}.

We apply a unitary transformation to block-diagonalize $O_L^T {\cal M}_d {\cal M}_d^\dagger O_L$ up to order ${\cal O}(m^3 \mu / M_D^2)$. At the low scale, this yields~\cite{Perez:2023zin}
\begin{widetext}
\begin{equation}\label{eq:M2}
\begin{split}
 \mathsf{M}^2 \equiv (V_L^\dagger O_L^T  {\cal M}_d {\cal M}_d^\dagger O_L V_L)^{ij} & \simeq \kappa^2_M\, \begin{pmatrix} m_e^2 \left(1 - \frac{|\mu_1|^2}{M_D^2}\right) & -  m_e  m_\mu \frac{\mu_1^*\mu_2}{M_D^2} & -  m_e  m_\tau \frac{\mu_1^* \mu_3}{M_D^2}\\
    -  m_\mu  m_e \frac{\mu_2^* \mu_1}{M_D^2} &  m_\mu^2 \left(1  - \frac{|\mu_2|^2}{M_D^2}\right) & -  m_\mu  m_\tau \frac{\mu_2^* \mu_3}{M_D^2} \\
    -  m_\tau  m_e \frac{\mu_3^* \mu_1}{M_D^2} & -  m_\tau  m_\mu \frac{\mu_3^* \mu_2}{M_D^2} &  m_\tau^2 \left( 1 - \frac{|\mu_3|^2}{M_D^2}\right) \end{pmatrix}~.
\end{split}
\end{equation}
\end{widetext} 
The decoupled eigenvalue, $ (V_L^T O_L^T  {\cal M}_d {\cal M}_d^\dagger O_L V_L)^{44}$, corresponds to the physical mass of the vector-like down quark, $M_D^2 \simeq \mu_1^2 + \mu_2^2 + \mu_3^2 + M_6^2$. The matrix $V_L$, 
\begin{equation}
V_L = \begin{pmatrix} \delta_{ij} & \mu_i^* m_i/ M_D^2 \\ - \mu_i m_i / M_D^2 & 1 \end{pmatrix}~,
\end{equation}
with $m_i$ being the masses of the charged leptons, satisfies $V_L V_L^\dagger = \mathbb{I} + {\cal O}(m^2 \mu^2 / M_D^4)$.

From the structure of $\mathsf{M}^2$ in Eq.~\eqref{eq:M2}, one obtains the general alignment
\begin{equation}
    m_d^2 \leq \kappa_M^2 m_e^2 \leq m_s^2 \leq \kappa_M^2 m_\mu^2 \leq m_b^2 \leq  \kappa_M^2 m_\tau^2~,
\end{equation} 
where $m_d$, $m_s$ and $m_b$ are the renormalizable physical down-type quark masses, corresponding to the square roots of the three eigenvalues of $\mathsf{M}^2$ (see Appendix~\ref{app:evaluebound} for a proof). Hence, the strange and bottom quark masses can be successfully reproduced from their charged-lepton partners after including renormalization-group running and mixing with the vector-like down quark. 

However, the experimental value for the down-quark mass satisfies $m_d(2 \text{ GeV}) \simeq 9 \, m_e$, which exceeds the maximal value attainable through fermion mixing alone. As discussed earlier, since quark-lepton unification is broken at a high scale, higher-dimensional operators of the form in Eq.~\eqref{eq:nonrensplitting} are  expected on effective-field-theory grounds to generate the splitting required between the down quark and electron masses.

The hermitian matrix in Eq.~\eqref{eq:M2} can be diagonalized by a unitary transformation, $\text{diag}(m_d^2,m_s^2,m_b^2) = \mathsf{U}_L^\dagger \mathsf{M}^2 \mathsf{U}_L$. The unitary matrix $\mathsf{U}_L$ then induces the physical CP-violating phase in the CKM matrix, 
\begin{equation}\label{eq:VCKM}
\mathsf{V}_{\rm CKM} \simeq \mathsf{O}_{uL}^T \mathsf{U}_L \mathsf{O}_L,
\end{equation}
where $\mathsf{O}_{uL}$ is defined through $\mathsf{O}_{uL}^T \mathsf{M}_u \mathsf{O}_{uR} = \text{diag}(m_u,m_c,m_t)$. Obtaining an ${\cal O}(1)$ CKM phase requires $\mu_i \sim M_D$~\cite{Vecchi:2014hpa,Valenti:2021rdu,Perez:2023zin} (see Appendix~\ref{app:diagonalizer} for details).
As a consequence, one has $M_6 \lesssim M_D$, and the quality bound in Eq.~\eqref{eq:boundquality} directly constrains the ratio $v_{\rm CP}^2/M_D$:
\begin{equation}\label{eq:qualityB}
 \left( \frac{\vCP}{10^9 \text{ GeV}}\right)^2 \!  \left(\frac{ 10^9 \text{ GeV}}{M_D}\right)
 \leq   3   \left(\frac{\Delta \bar \theta_{\rm QCD}}{10^{-10}}\right)~,
\end{equation}
where we have taken the Pati-Salam cutoff scale at $M_{\rm Pl}$.

\section{Baryon number violation}
\label{sec:BNV}
In this theory, there are several processes that mediate baryon number violation in one unit. However, we expect a distinctive dominant baryon-number-violating mode, $n \to K^+ e^-$, to be observable in current or near-future nucleon decay and neutrino experiments. This decay therefore constitutes a smoking-gun signal of this theory.

Let us begin by considering the scalar representations 
\begin{equation}
    \Phi_4^{a,b} \sim (4,1,1/2)_{\rm PS} = \begin{pmatrix} \phi_4^\alpha \\ \phi_4^0\end{pmatrix}~.
\end{equation}
The electromagnetically neutral components $\phi_4^0$ are responsible for the spontaneous CP breaking, while the colored fields $\phi_4^\alpha$, with $\alpha=1,2,3$ being a $\SU(3)_C$ index, are scalar leptoquarks. These fields mediate $\Delta B = 1$ baryon-number-violating interactions with the SM fermions through the interactions
\begin{equation}\label{eq:YukBNV}
\begin{split}
-{\cal L} &\supset \frac{\lambda_{a}^{i}}{\sqrt{2}} \left( \epsilon_{\alpha \beta \gamma} (\overline{D^c_L})^\alpha  d_{Ri}^\beta \, \phi_{4a}^\gamma - \bar e_{Ri} D_L^\alpha \phi_{4a \alpha}^*\right) \\
&\qquad + (a \leftrightarrow b) + {\rm h.c.},
\end{split}
\end{equation}
which originate from the Yukawa terms in Eq.~\eqref{eq:F6Yukawa}. Throughout, Greek indices $\alpha,\beta,\gamma$ denote $\SU(3)_C$ color indices, Roman indices $i,j,k, \ell,m,n$ label fermion generations, and the labels $a, b$ distinguish the two scalar fields. 

Integrating out the scalar leptoquarks and rotating the fermions to their mass eigenbasis, we obtain the effective operator
\begin{equation}\label{eq:operator7}
 {\cal H}_{\rm eff} \supset -\frac{C_{ijk\ell,s}}{\Lambda^2_{\rm GUT}}   (d^\alpha_i, d^\beta_j)_R \, \bar e_k P_L d_\ell^\gamma \, \epsilon_{\alpha \beta \gamma} + \text{h.c.}~,
\end{equation}
where $(x,y)_{R,L} \equiv x^T C P_{R,L} y$ with $C$ the charge-conjugation matrix and $P_{R,L}$ the chiral projectors. Since $(x^\alpha,x^\beta)_{R,L}$ is symmetric in the color indices, its contraction with the totally antisymmetric tensor $\epsilon_{\alpha \beta \gamma}$ vanishes unless different flavors are involved. Consequently, strangeness changing baryon number violation is expected. Taking  $i,j=1,2$ with $i\neq j$ and $\ell=1$ induces the neutron two-body decay channels $n \to K^+ (e^-,\mu^-)$.

Above, all Yukawa couplings and mixing parameters have been collected into the c-number
\begin{equation}\label{eq:Cfactor}
C_{ijk\ell,s} \equiv \sum_{m,n} \xi_{ijk \ell m n} \,  \lambda_{s}^m \lambda_s^n \frac{\Lambda_{\rm GUT}^2}{2M_{\Phi_4}^2}~,
\end{equation}
where $s=a,b$ labels the two physical scalar fields\footnote{In general, these two fields mix via terms in the scalar potential. Then, the Yukawa couplings in Eq.~\eqref{eq:Cfactor} are mapped to those appearing in Eq.~\eqref{eq:YukBNV} by a rotation.}.
The normalization scale $\Lambda_{\rm GUT} \equiv 10^{15}$ GeV is introduced as a convenient benchmark to render Eq.~\eqref{eq:Cfactor} dimensionless. The dependence on the fermion mixing matrices is encoded in the parameter $\xi_{ijk\ell mn}$, defined as
\begin{equation}\label{eq:xiparameters}
\xi_{ijk\ell m n} \equiv \tilde V_R^{4i} \tilde V_R^{nj} \mathsf{O}_R^{mk} (\tilde V_L)^{4\ell}.
\end{equation}
The unitary matrices $\tilde V_{L,R}$ diagonalize the down-type quark mass matrix in Eq.~\eqref{eq:Md}, 
\begin{equation}
    \tilde V_L^\dagger {\cal M}_d \tilde V_R = \text{diag}(m_d, m_s,m_b, M_D)~.
\end{equation}
Mapping this expression onto the results of Sec.~\ref{sec:masses}, we identify $\tilde V_L \equiv O_L V_L U_L$, which implies
\begin{equation}
    \tilde V_L^{4\ell} = -\sum_{j} \mathsf{U}_L^{j \ell} \frac{\mu_j m_j}{ M_D^2} \simeq -\frac{\mu_\ell m_\ell}{M_D^2}~,
\end{equation}
where $m_i$ are the charged-lepton masses. In the last step, we have used the strong hierarchy $m_e \ll m_\mu \ll m_\tau$, which implies that $\mathsf{U}_L^{j\ell}\sim \delta^{j\ell}$ up to small corrections of order ${\cal O}(m_e/m_\mu)$, ${\cal O}(m_\mu/m_\tau)$ or ${\cal O}(m_e/m_\tau)$. We refer the reader to Appendix~\ref{app:diagonalizer} for an explicit proof.

While the left-handed rotation $\tilde V_L$ contains light-heavy mixing suppressed as $\sim m_i \mu_i /M_D^2$, the right-handed rotation $\tilde V_R$ generically features ${\cal O}(1)$ entries. This can be understood as follows. 
Let us write,
\begin{equation}\label{eq:RH}
    {\cal M}_d^\dagger {\cal M}_d = D^2 + R R^\dagger,
\end{equation}
where $D^2 = \kappa_M^2 \, \text{diag}(m_e^2,m_\mu^2,m_\tau^2,0)$, and the heavy direction is defined by the vector $R^T=(\mu_1,\mu_2,\mu_3,M_6)$. In the limit $m_i \to 0$, this matrix has a single nonzero eigenvalue $M_D^2 = R^T R$ and a three-dimensional kernel. Since $\mu_i \sim M_6$, the eigenvector $R^T$ is a sizable admixture of the original right-handed states ($d_R^i$,$D_R$), so the fourth column of $V_R$ is manifestly unsuppressed. The three remaining eigenvectors, once the small diagonal perturbation $D^2$ lifts their degeneracy, are generically also not aligned with the flavor basis. Therefore, $\tilde V_R$ is expected to contain order-one entries throughout.

Thus, taking into account the above considerations, and using also that $\mu_i \sim M_D$, we estimate 
\begin{equation}\label{eq:xiorder}
{\cal O}(\xi_\ell) = \frac{m_\ell}{M_D}~.
\end{equation}

The operator in Eq.~\eqref{eq:operator7} is rather peculiar. Although it naively appears to be of dimension six, it mediates nucleon decay into a lepton; however, it is well known that dimension-six baryon-number violating operators preserve $B-L$~\cite{Weinberg:1979sa,Wilczek:1979hc}. Indeed, the effective interaction in Eq.~\eqref{eq:operator7} is secretly dimension seven, arising from integrating out both the scalar leptoquark $\phi_4^\alpha$ and the vector-like down quark. The additional heavy scale $M_D$ is implicitly encoded in the mixing factors appearing in Eq.~\eqref{eq:xiparameters}. Operators mediating $\Delta B = 1$ and $\Delta (B-L)=2$ transitions have been studied extensively in the literature~\cite{Weinberg:1980bf,Weldon:1980gi,Aoki:2017puj,Heeck:2019kgr}, and can arise, for example, in $\mathrm{SO}(10)$ models with $B-L$ breaking~\cite{Babu:2012iv}. In generic UV completions, one would also expect the presence of lower-dimensional, $B-L$-preserving operators. By contrast, in theories of quark-lepton unification, and in particular in this Pati-Salam realization of the Nelson-Barr mechanism, a single class of decay channels, $n \to K^+ (e,\mu)^-$, is expected to dominate, providing a distinctive smoking-gun signal of the theory. All other baryon-number-violating modes are parametrically suppressed.

The neutron decay rate for this channel is given by~\cite{Nath:2006ut}
\begin{equation}
\Gamma = \frac{m_n}{32\pi} \left( \! 1 - \frac{m_K^2}{m_n^2}\! \right)^2  \!\! |\langle K^+|(d s)_R d_L|n\rangle|^2 \frac{A^2|C|^2}{\Lambda_{\rm GUT}^4}~.
\end{equation}
The factor $A$ accounts for the renormalization-group running of the Wilson coefficient of the operator in Eq.~\eqref{eq:operator7} under QCD, 
\begin{equation*}
    A \sim  \bigg(\frac{\alpha_s(\mu_H)}{\alpha_s(m_b)}\bigg)^{\tfrac{6}{25}} \! \bigg(\frac{\alpha_s(m_b)}{\alpha_s(m_t)}\bigg)^{\tfrac{6}{23}}\! \bigg (\frac{\alpha_s(m_t)}{\alpha_s(\vCP)}\bigg)^{\tfrac{2}{7}} \!\!\! \sim 1.7~,
\end{equation*}
evolving from the UV scale $\vCP$ down to the hadronic scale $\mu_H \sim 2$ GeV, where matrix elements are evaluated on the lattice. Under isospin symmetry, the matrix element satisfies $\langle K^+| (d s)_R d_L | n\rangle = -\langle K^0|(u s)_R u_L|p\rangle$~\cite{Aoki:2013yxa}, which has been computed on the lattice and is given by~\cite{Aoki:2017puj}
\begin{equation}
   \langle K^0|(u s)_R u_L|p\rangle = 0.103(3)(11)\text{ GeV}^2~. 
\end{equation} 
As a crude estimate, 
\begin{equation}\label{eq:estimate}
|C|^2 \sim \frac{M_D^2 m_e^2}{M_{\Phi_4}^4\vCP^4} \Lambda_{\rm GUT}^4 \sim \frac{M_D^2 m_e^2}{\vCP^8} \Lambda_{\rm GUT}^4 ,
\end{equation}
where we have taken the Yukawa couplings in Eq.~\eqref{eq:Cfactor} to be of order $\lambda \sim M_D / \vCP$, since $\mu \sim M_D$ is required to generate an ${\cal O}(1)$ CKM phase. Moreover, $\xi \sim m_e / M_D$ according to Eq.~\eqref{eq:xiorder}, because $\ell=1$. In the final step of Eq.~\eqref{eq:estimate}, we have used that $M_{\Phi_4} \sim \vCP$, as expected for a generic scalar potential (see Appendix~\ref{app:potential} for an explicit discussion).

\begin{figure}
 \centering
    \includegraphics[width=1\columnwidth]{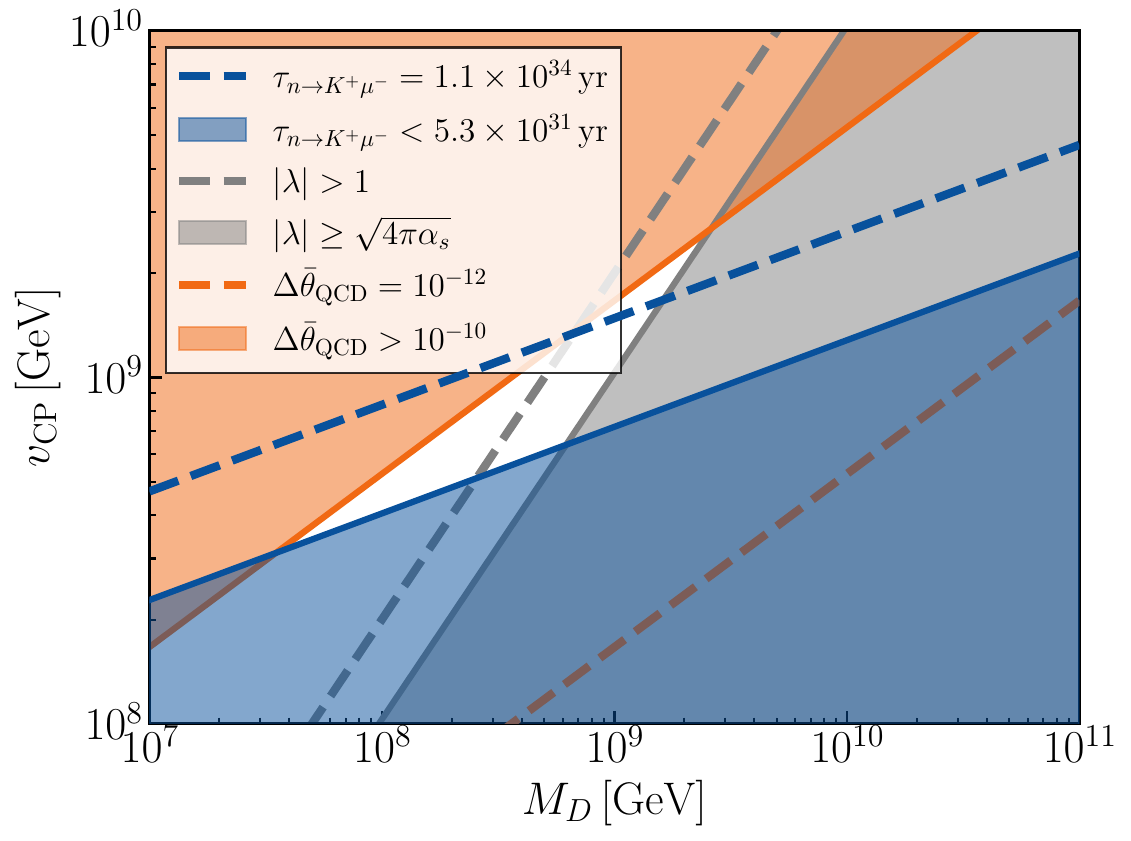}
    \caption{Allowed parameter space in the theory. The horizontal axis shows the mass of the vector-like down quark $M_D$, while the vertical axis displays the scale of spontaneous CP violation, $\vCP$. The blue-shaded region is excluded by the experimental bound on the partial lifetime of the decay mode $n \to K^+ \mu^-$, $\tau/\text{Br}(n \to K^+ \mu^-) > 3.2 \times10^{31}$ years~\cite{Frejus:1991ben}, corresponding to Eq.~\eqref{eq:BNVbound}. The orange-shaded region is excluded by the quality of the Nelson-Barr mechanism, obtained by requiring $\Delta \bar \theta_{\rm QCD} < 10^{-10}$, which leads to the bound in Eq.~\eqref{eq:boundquality}. The gray-shaded region denotes parameter values that violate the perturbative benchmark in Eq.~\eqref{eq:pert}. In this region, the Yukawa coupling $\lambda$ is driven towards larger values in the UV and may eventually develop a Landau pole. The dashed line shows the more conservative benchmark $|\lambda| = 1$.
    Also shown are projected sensitivities. The dashed blue line corresponds to the expected reach of  $\tau/\text{Br}(n\to K^+ \mu^-) > 1.1\times10^{34}$ years from DUNE~\cite{Alt:2020rii,DUNE:2024wvj}, while the dashed orange line indicates the anticipated sensitivity to  $\bar \theta_{\rm QCD}$ from upcoming hadron-EDM experiments~\cite{n2EDM:2021yah,EuropwanEDMprojects:2025okn,nEDM:2019qgk,TUCAN:2022koi}. As the plot shows, the currently allowed parameter space will be probed by the next generation of nucleon decay and EDM experiments in the foreseeable future.}
    \label{fig:plot}
\end{figure}

In Fig.~\ref{fig:plot} we estimate the viable parameter space of this theory in the $M_D$ vs $\vCP$ plane. The constraints included in the figure are described below.

The current lower bounds on the partial mean lifetimes of the decay modes $n \to K^+ (e,\mu)^-$, are set by the Fr\'ejus experiment~\cite{Frejus:1991ben}, yielding  $\tau / \text{Br}(n \to K^+e^-)> 3.2 \times 10^{31} \text{ years}$ and $\tau /\text{Br}(n \to K^+\mu^-)> 5.7 \times 10^{31}$ years at $90\%$ confidence level (C.L.).
Such decays constrain the ratio $\vCP^4/M_D$ to satisfy
\begin{equation}\label{eq:BNVbound}
\left(\frac{\vCP}{10^9\text{ GeV}}\right)^4 \left(\frac{10^9 \text{ GeV}}{M_D}\right)  > 0.34~.
\end{equation}
Parameter values that violate this bound correspond to an excessively fast nucleon decay rate and are therefore excluded. The excluded region is shown in  blue in Fig.~\ref{fig:plot}.

The limits on the decay modes $n \to K^+ (e^-,\mu^-)$ are a couple of orders of magnitude weaker than the leading bounds on nucleon decay~\cite{Super-Kamiokande:2020wjk}, and have not been updated for more than three decades. This situation is partly due to the experimental challenges associated with these modes, as discussed below. In addition, experimental programs have historically focused on the so-called {\it normal} $\Delta B = \Delta L$ decay modes~\cite{Weinberg:1980bf}, such as $p \to e^+ \pi^0$, which arise in minimal grand unified theories such as $\SU(5)$.

Super-Kamiokande (SK) does not have a published bound on $n\to K^+\ell^-$, with $\ell=e,\mu$. In these two-body neutron decays, the kaon momentum is below the Cherenkov threshold in water, up to nuclear smearing effects. The $K^+$ is therefore not observed directly; instead, the event must be reconstructed from its decay products together with the accompanying charged lepton. Although one might expect SK to be sensitive to $n\to K^+e^-$ at a level comparable to its searches for $p\to \bar\nu K^+$~\cite{Super-Kamiokande:2014otb}, the two modes have rather different experimental systematics. The latter benefits from a well-developed kaon-decay tag, including the monoenergetic muon from $K^+\to\mu^+\nu$ and, in oxygen, prompt nuclear de-excitation $\gamma$'s. By contrast, $n\to K^+e^-$ contains an additional prompt charged lepton. Depending on the $K^+$ decay mode, this can lead to a two-ring or three-ring topology, often with overlapping e-like rings, making event reconstruction and the rejection of atmospheric-neutrino backgrounds considerably more challenging than in canonical clean modes. A dedicated SK analysis would therefore be needed to determine whether a comparable sensitivity can be achieved, and could potentially improve substantially on the existing Fr\'ejus bound.

Future sensitivity is expected to improve, however, at Hyper-Kamiokande~\cite{Hyper-Kamiokande:2018ofw}, primarily because of its much larger exposure compared to Super-Kamiokande, together with improved reconstruction and tagging of kaon decay products. Nevertheless, Hyper-Kamiokande remains a water Cherenkov detector, so the sub-Cherenkov $K^+$ poses the same basic topological challenge as in Super-Kamiokande. By contrast, DUNE~\cite{DUNE:2020ypp} benefits from liquid-argon imaging, which allows for a more detailed reconstruction of the $K^+$ and its decay chain. JUNO~\cite{JUNO:2021vlw}, on the other hand, should also be sensitive to kaon modes because liquid scintillator can exploit the delayed $K^+$ decay signature.

In Fig.~\ref{fig:plot}, we show as a dashed blue line the projected DUNE sensitivity, $\tau/\mathrm{Br}(n \to e^- K^+) > 1.1 \times 10^{34}\ {\rm yr}$, obtained assuming an exposure of 400 kiloton$\cdot$yr at $90\%$ C.L.~\cite{Alt:2020rii,DUNE:2024wvj}. There is not yet an official dedicated $n \to K^+ e^-$ projection in the literature for JUNO. However, scaling from the projected sensitivity to $p \to \bar\nu K^+$~\cite{JUNO:2022qgr} suggests a possible reach around $10^{34}$ years, subject to a dedicated efficiency and background study. Hyper-Kamiokande quotes a projected sensitivity of order $10^{34}$ years for this channel~\cite{Hyper-Kamiokande:2018ofw}. This projection should be interpreted with some caution: if the search is limited by multi-ring atmospheric-neutrino backgrounds, the improvement relative to a possible SK analysis may be smaller than suggested by exposure scaling alone.

Another constraint in the $M_D$-$\vCP$ plane arises from requiring the Yukawa couplings to remain perturbative. The leading contributions to the running of $\lambda$ come from QCD corrections and, when sizable, from the $\lambda$ couplings themselves, leading to
\begin{equation}\label{eq:RGEyuk}
\frac{\text{d} \lambda}{\text{d} \ln \mu} \simeq \frac{1}{2\pi^2} \lambda \big(\lambda^\dagger \lambda  - 4 \pi \alpha_s  \big),
\end{equation}
where we have assumed $|\lambda_a|\sim |\lambda_b|\equiv |\lambda|$. The same leading behavior applies to the color components $\phi_4^\alpha$, which mediate the diquark and leptoquark interactions, and to the neutral component $\phi_4^0$, which generates the bridge mass. As a perturbative benchmark, we demand that the Yukawa self-interaction does not overcome the QCD screening term at low energies, so that the norm of the coupling vector is not driven towards larger values in the UV. From Eq.~\eqref{eq:RGEyuk}, this implies
$|\lambda| < \sqrt{4\pi \alpha_s}$ at low energies, which in turn gives
\begin{equation}\label{eq:pert}
    M_D < \sqrt{\pi \alpha_s} \, \vCP.
\end{equation}
 The region of parameter space that does not satisfy the condition in Eq.~\eqref{eq:pert} is shaded in gray.
We also show the more conservative benchmark $|\lambda| =1$ as a dashed line.

The parameter space of the theory is also constrained by the quality of the Nelson-Barr mechanism, which imposes an upper bound on the ratio $\vCP^2/M_D$ by demanding $\Delta \bar \theta_{\rm QCD} < 10^{-10}$, as given in Eq.~\eqref{eq:qualityB}. The region excluded by this requirement is shown in orange in Fig.~\ref{fig:plot}. 

In this context, a sensitivity improvement of roughly two orders of magnitude in $\bar \theta_{\rm QCD}$ is anticipated in upcoming hadron EDM experiments~\cite{EuropwanEDMprojects:2025okn}.
We indicate this reach by an orange dashed line in Fig.~\ref{fig:plot}, corresponding to the projected sensitivity of next-generation experiments such as n2EDM at PSI~\cite{n2EDM:2021yah,nEDM:2019qgk}, TUCAN at TRIUMF~\cite{TUCAN:2022koi}, pEDM at BNL~\cite{pEDM:2022ytu}, and cpEDM at CERN~\cite{CPEDM:2019nwp}, which aim to probe values as small as $\bar \theta_{\rm QCD} < 10^{-12}$.

Remarkably, the combined reach of near-future nucleon decay and hadronic EDM experiments will be sufficient to decisively test this minimal Pati-Salam realization of the Nelson-Barr mechanism.

Additional baryon-number non-conserving processes are induced by the presence of the fermionic multiplet $F_6$, and also by the scalar $\Phi_{10}$. 
In particular, the kinetic term of the antisymmetric fermionic representation contains a diquark coupling involving the vector leptoquarks $X_\mu^\alpha \sim(3,1,2/3)_{\rm SM}$. This precludes assigning a consistent baryon number to $X_\mu$, as illustrated by the following interactions:
\begin{equation}
  \! {\cal L} \! \supset \!  \frac{g_4}{\sqrt{2}}\left( \! \epsilon_{\alpha \beta \gamma} \overline{(D^c)_L^\alpha} \gamma^\mu D_{L}^\beta + \overline{d_{\gamma L}} \gamma^\mu e_L \! \right) \! X_\mu^\gamma+ \text{h.c.}.
\end{equation}
The exchange of $X_\mu^\alpha$ therefore mediates $\Delta B= 1$ processes. However, these effects are strongly suppressed.
First, the vector leptoquark mass is expected to be $M_X \simeq \sqrt{2\pi \alpha_s} \, \Lambda_{\rm QL} \sim 10^{14} \text{ GeV}$, which is already close to the most stringent bounds from baryon number violation. Second, mixing between the vector-like down quark and the SM down-type quarks further suppresses the decay rate by at least a factor $(m_\mu / M_D)^2 \ll  {\cal O}(10^{-8})$, even when saturating current lower bounds on vector-like quark masses~\cite{Alves:2023ufm}. Therefore, no observable signal generated from these processes is expected in current or foreseeable nucleon decay experiments. 

Turning to the scalar $\Phi_{10} \sim (10,1,1)_{\rm PS}$, besides the SM singlet responsible for breaking $\mathrm{U}(1)_{B-L} \otimes \mathrm{U}(1)_R \to \mathrm{U}(1)_Y$, this representation contains a scalar diquark $\phi^{\alpha \beta}_{10} \sim (6,1,4/3)_{\rm SM}$ and a scalar leptoquark $\phi_{10}^\alpha \sim (3,1,2/3)_{\rm SM}$, see Eq.~\eqref{eq:scalars}. 
In addition to generating Majorana masses for the right-handed neutrinos, the Yukawa interaction $\mathsf{Y}_{\nu R} F_u^T C F_u \Phi_{10}^*$ also induces
\begin{equation}
\!\! -{\cal L} \supset \mathsf{Y}_{\nu R} \bigg[ (u^\alpha \!\!, u^\beta)_R \, \phi_{10 \alpha \beta}^* +   (u^\alpha \!\! , \nu)_R \, \phi_{10\alpha }^* \bigg] \! +\! {\rm h.c.}.
\end{equation}
Although $\phi_{10}^\alpha$ participates only in purely leptoquark interactions, scalar potential terms such as those shown in Eq.~\eqref{eq:mix410} induce mixing between $\phi_{10}^{\alpha}$ and $\phi_4^\alpha$.  Due to the hierarchy between their vevs, $v_4 \sim \vCP \ll v_{10}$, the resulting mixing angle is suppressed by $\theta_\Phi  \lesssim \vCP/v_{10}$, as shown in Appendix~\ref{app:potential}. On top of this, the high-scale type-I seesaw with its Dirac scale fixed by the top mass implies a tiny mixing angle between right-handed and the active neutrinos, which further suppresses $\Delta B = 1$ processes involving neutrinos in the final states. Consequently, no additional decay channels are expected to compete with the $n \to K^+ (e,\mu)^-$ mode discussed above.

\section{The hierarchy of scales}
\label{sec:Scales}

A characteristic feature of this theory is that the scale of spontaneous CP violation, $v_{\rm CP}$, and the mass of the new vector-like down quark, $M_D$, are restricted to a relatively narrow window of roughly two orders of magnitude around $10^9~$GeV, as one reads from Fig.~\ref{fig:plot}. This follows from the interplay of two independent constraints. On the one hand, the quality of the Nelson-Barr mechanism imposes an upper bound on the ratio $v_{\rm CP}^2/M_D$, see Eq.~\eqref{eq:qualityB}. On the other hand, the non-observation of baryon-number violation sets a lower bound on the ratio $v_{\rm CP}^4/M_D$, as given in Eq.~\eqref{eq:BNVbound}.

The resulting separation from the electroweak scale is substantial and, as in many UV extensions involving high mass scales, is accompanied by the usual naturalness concerns.\footnote{Indeed, even the simplest theories for quark-lepton unification at the low scale already face naturalness issues~\cite{FileviezPerez:2023rxn}.}

What about the remaining scales of the theory? In principle, the scale of quark-lepton unification could be lowered by modifying the neutrino sector. Following Ref.~\cite{FileviezPerez:2013zmv}, one may introduce a pair of gauge-singlet fermions, $N$, leading to the interactions
\begin{equation}
   - {\cal L} \supset \mu N^T C N +  \lambda_R \bar F_{u\nu} N \Phi_4 + \text{h.c.}~,
\end{equation}
which generate active-neutrino masses via the inverse seesaw mechanism~\cite{Mohapatra:1986aw,Mohapatra:1986bd}
\begin{equation}
    m_\nu \sim \mu \, m_t^2/ ( \lambda_R \, \vCP)^2~.
\end{equation}
This construction would, in principle, allow for a lower quark-lepton unification scale for an arbitrarily small $\mu$.

However, independently of the neutrino mass generation mechanism, requiring realistic down-quark masses from non-renormalizable operators sets the ratio of the quark-lepton unification scale to the Pati-Salam cut-off scale, $\Lambda_{\rm QL}/M_{\rm Pl} \sim y_d = {\cal O}(10^{-5})$. This in turn fixes the quark-lepton unification scale to be $\Lambda_{\rm QL} = {\cal O}(10^{14})$ GeV, in agreement with the scale required by the canonical type-I seesaw mechanism. The theory therefore exhibits a strongly hierarchical pattern of scales, as summarized schematically in the left column of Fig.~\ref{fig:summary}.

\begin{figure}[t]
\centering
\begin{minipage}{0.49\columnwidth}
\centering
\resizebox{\linewidth}{!}{%
\begin{tikzpicture}[
  scaleBox/.style={draw, rounded corners, thick, align=left, inner sep=6pt, fill=gray!5},
  tick/.style={thick},
  lab/.style={font=\small},
  >={Stealth[length=2mm]}
]
\node[lab,right=6pt] at (0.5,5.5) {{\bf High scale}};
\draw[thick, -{Stealth}] (0,0) -- (0,5) node[right] {$\,$Energy [GeV]};

\draw[tick] (-0.18,4) -- (0.1,4);
\node[lab,left=6pt] at (-0.18,4) {$10^{14}$};

\draw[tick] (-0.18,2.5) -- (0.1,2.5);
\node[lab,left=6pt] at (-0.18,2.5) {$10^9$};

\draw[tick] (-0.18,1) -- (0.1,1);
\node[lab,left=6pt] at (-0.18,1) {$10^2$};

\draw[fill=internationalorange!15, draw=white, thick, rounded corners]
  (0.1,2.3) rectangle (3,2.8);
\node[lab,right=6pt] at (0,2.55) {$\vCP$};

\draw[fill=internationalorange!15, draw=white, thick, rounded corners]
  (0.1,3.9) rectangle (3,4.1);
\node[lab,right=6pt] at (0,4.05) {$\Lambda_{\rm QL}$};

\draw[fill=internationalorange!15, draw=white, thick, rounded corners]
  (0.1,0.8) rectangle (3,1.05);
\node[lab,right=6pt] at (0,0.91) {$M_Z$};

\node[scaleBox, right=10pt] at (0.75,4) {$\Phi_{15}, \Phi_{10}, X_\mu$};
\node[scaleBox, right=10pt] at (0.75,2.5) {$D\subset F_6$, $\Phi_4$};
\node[scaleBox, right=10pt] at (0.75,0.5) {SM fields};

\end{tikzpicture}%
}
\end{minipage}
\hfill
\begin{minipage}{0.45\columnwidth}
\centering
\resizebox{\linewidth}{!}{%
\begin{tikzpicture}[
  scaleBox/.style={draw, rounded corners, thick, align=left, inner sep=6pt, fill=gray!5},
  tick/.style={thick},
  lab/.style={font=\small},
  >={Stealth[length=2mm]}
]

\node[lab,right=6pt] at (0.5,5.5) {{\bf Low scale}};
\draw[thick, -{Stealth}] (0,0) -- (0,5) node[right] {$\,$Energy [GeV]};

\draw[tick] (-0.18,4.3) -- (0.1,4.3);
\node[lab,left=6pt] at (-0.18,4.3) {$10^{15}$};

\draw[tick] (-0.18,2.5) -- (0.1,2.5);
\node[lab,left=6pt] at (-0.18,2.5) {$10^9$};

\draw[tick] (-0.18,3.2) -- (0.1,3.2);
\node[lab,left=6pt] at (-0.18,3.2) {$10^{11}$};

\draw[tick] (-0.18,1) -- (0.1,1);
\node[lab,left=6pt] at (-0.18,1) {$10^2$};

\draw[fill=internationalorange!15, draw=white, thick, rounded corners]
  (0.1,2.3) rectangle (3,2.8);
\node[lab,right=6pt] at (0,2.55) {$\vCP$};

\draw[fill=internationalorange!15, draw=white, thick, rounded corners]
  (0.1,4.5) rectangle (3,4.8);
\node[lab,right=6pt] at (0,4.65) {$\Lambda_{\rm GUT}$};

\draw[fill=internationalorange!15, draw=white, thick, rounded corners]
  (0.1,0.8) rectangle (3,1.05);
\node[lab,right=6pt] at (0,0.91) {$M_Z$};

\draw[fill=internationalorange!15, draw=white, thick, rounded corners]
  (0.1,3.3) rectangle (3,3.5);
\node[lab,right=6pt] at (0,3.4) {$\Lambda_{\rm QL}$};

\node[scaleBox, right=10pt] at (0.75,2.5) {$D\subset F_6$, $\Phi_4$};
\node[scaleBox, right=10pt] at (0.75,3.4) {$\Phi_{15}, X_\mu$};
\node[scaleBox, right=10pt] at (0.75,0.5) {SM fields};

\end{tikzpicture}%
} 
\end{minipage} 
\caption{Schematic distribution of the field masses and relevant energy scales of the theory. The three main scales of the theory are shown: the electroweak scale ($M_Z$), the scale of spontaneous CP violation ($\vCP \sim \langle \phi_4^0 \rangle$), and the quark-lepton unification scale ($\Lambda_{\rm QL} \sim \langle \phi_{15}^0 \rangle$).  The matter content consists of the SM fields, embedded in three copies of $F_{QL} \sim (4,2,0)_{\rm PS}$, $F_{u\nu} \sim (4,1,1/2)_{\rm PS}$, and $F_{de} \sim (4,1,-1/2)_{\rm PS}$, together with the Higgs doublet $H \sim (1,2,1/2)_{\rm PS}$. The theory also contains a vector-like down quark $D \subset F_6 \sim (6,1,0)_{\rm PS}$ responsible for (i) correcting the $\mathsf{M}_d= \mathsf{M}_e$ relation for the two heaviest generations and (ii) solving the strong CP problem {\it \`a la} Nelson-Barr. The scalar sector includes $\Phi_{15}$, responsible for breaking $\SU(4)_C \to \SU(3)_C \otimes \mathrm{U}(1)_{B-L}$, and $\Phi_4$, whose vev triggers spontaneous CP violation. In the high-scale $\Lambda_{\rm QL}$ scenario (left panel), $\Phi_{10}$ breaks $\mathrm{U}(1)_{B-L} \otimes \mathrm{U}(1)_R \to \mathrm{U}(1)_Y$ with
$\langle\phi_{10}^0\rangle$ of order the quark-lepton unification scale. Then, the hierarchy between the quark-lepton unification scale and the electroweak scale ($\Lambda_{\rm QL} / M_Z  \sim 10^{12}$) generates active neutrino masses via the type-I seesaw mechanism, consistently with neutrino data. In turn, the gap between the quark-lepton unification scale and the Planck scale ($\Lambda_{\rm QL}/M_{\rm Pl}\sim 10^{-5}$) accounts for the observed down-quark mass through non-renormalizable operators. In the low-scale $\Lambda_{\rm QL}$ scenario (right panel), $v_{\rm CP} \ll \Lambda_{\rm QL} ={\cal O}(10^{11})$ GeV, which is consistent with $\text{SO}(10)$ grand unification at $\Lambda_{\rm GUT} \gtrsim 10^{15}$ GeV, and consistent down quark masses if these are generated by physics mediated at the GUT scale. In this scenario, neutrinos are expected to get mass via the inverse seesaw mechanism.}
\label{fig:summary}
\end{figure}

Alternatively, a lower quark-lepton unification scale could be accommodated if the non-renormalizable operators generating the down-quark masses are suppressed by a cutoff below the Planck scale. Indeed, we may ask whether ${\cal G}_{\rm QL}$ can be embedded in a Grand Unified Theory (GUT). In this case, the cutoff of the effective quark-lepton unification theory should be interpreted as the GUT scale, $\Lambda_{\rm PS}\simeq \Lambda_{\rm GUT}$. This remains compatible with the Nelson-Barr construction provided that the higher-dimensional operators responsible for inducing a shift in $\bar \theta_{\rm QCD}$, see Eq.~\eqref{eq:dim5op}, are still generated only at the Planck scale.

The result obtained by running the gauge couplings of the Standard Model below $\Lambda_{\rm QL}$, and the gauge couplings of ${\cal G}_{\rm QL}$ above $\Lambda_{\rm QL}$, for the benchmark scales $\vCP = 10^9$ GeV and $\Lambda_{\rm QL} = v_{10}= 10^{14}$ GeV, is shown in the upper panel of Fig.~\ref{fig:GUT}. See Appendix~\ref{app:RGEs} for details. The running suggests that gauge coupling unification is approached but not achieved. This situation is analogous to the familiar case of minimal $\SU(5)$~\cite{Georgi:1974sy}, and could be remedied by threshold corrections, for example through mass splittings among the scalar multiplets involved in the breaking, or among additional states belonging to larger $\SO(10)$ representations.

Nevertheless, since in the gauge-coupling unification picture the GUT suppressed operators 
\begin{equation}
\frac{1}{\Lambda_{\rm GUT}} \bar F_{QL} H \Phi_{15} F_{de}  + \text{h.c.}~,
\end{equation}
can generate the down-quark mass, and neutrinos can get mass via the inverse seesaw mechanism explained above, the requirement of a high scale $\Lambda_{\rm QL}$ can be relaxed.

Given that $v_{\rm CP}$ is fixed by the baryon-number and Nelson-Barr
quality constraints, $\Lambda_{\rm QL}$ is then determined by requiring gauge-coupling unification. As shown in Fig.~\ref{fig:GUT}, we find $\Lambda_{\rm QL}=2\times10^{11}$ GeV, and $\Lambda_{\rm GUT}=3\times10^{15}$ GeV.
These scales are also consistent with the requirement that
$\langle\Phi_{15}\rangle/\Lambda_{\rm GUT}$ reproduces the down-quark
Yukawa coupling.
The resulting unification scale is close to the range probed by
baryon-number-violating processes, in particular proton decay, and may be in mild tension with current bounds depending on the heavy gauge-boson spectrum and the details of the GUT UV completion. This conclusion is sensitive to threshold effects: mass splittings among the heavy UV states can shift the apparent unification scale and may relax the proton-decay constraints. The alternative hierarchy of scales is illustrated in the right column of Fig.~\ref{fig:summary}.

\begin{figure}
 \centering
    \includegraphics[width=1\columnwidth]{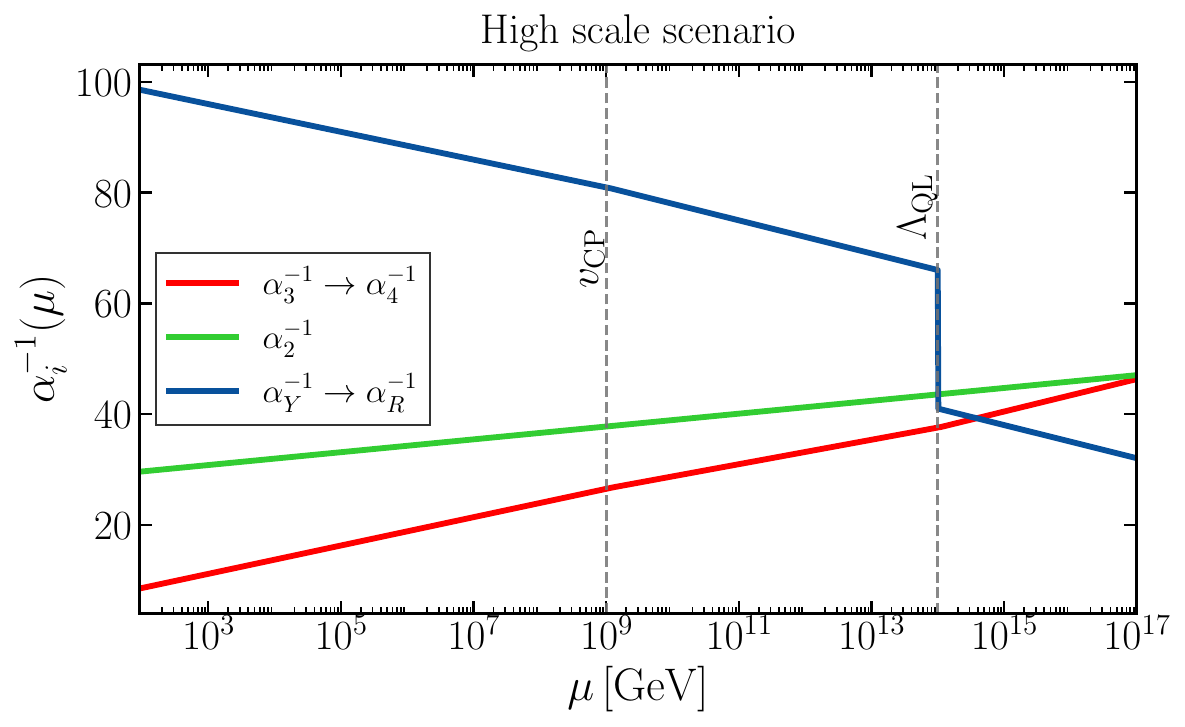}
    \includegraphics[width=1\columnwidth]{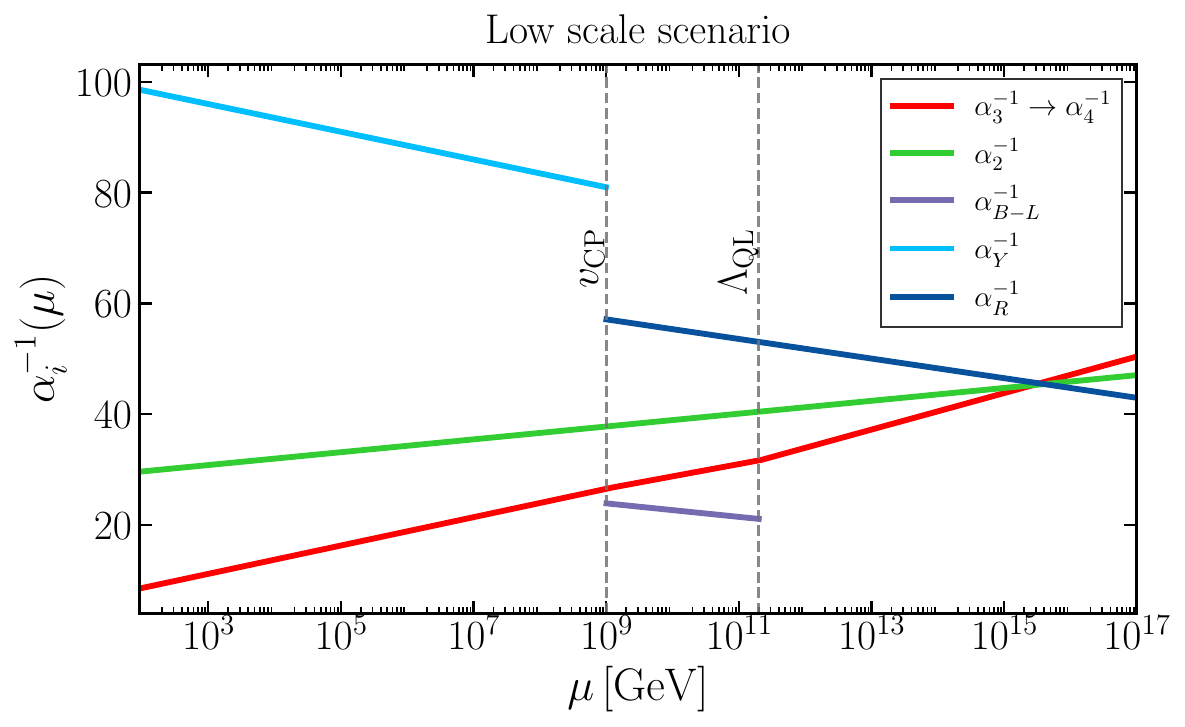}
    \caption{One-loop running of the gauge couplings in the two quark-lepton-unification scenarios. The upper panel corresponds to high-scale quark-lepton unification, $\Lambda_{\rm QL}=10^{14}$ GeV, while the lower panel corresponds to low-scale quark-lepton unification, $\Lambda_{\rm QL}=2\times10^{11}$ GeV. The lower-panel running points toward an $\SO(10)$-like unification scale $\Lambda_{\rm GUT}\simeq 3\times10^{15}$ GeV. Details of the gauge-coupling RGEs are given in Appendix~\ref{app:RGEs}.}
    \label{fig:GUT}
 \end{figure}
In addition to the usual naturalness problem of the SM Higgs, which we do not attempt to resolve here, the gap between the two high scales of the theory, $\vCP$ and either a high $\Lambda_{\rm LQ}$ or $\Lambda_{\rm GUT}$, introduces further sources of tuning. In the high-scale quark-lepton unification scenario, as discussed in Appendix~\ref{app:potential}, components of the scalar fields $\Phi_4$ and $\Phi_{10}$ with identical SM quantum numbers mix through the scalar potential, thereby linking the CP-breaking scale to the quark-lepton unification scale. Higher-dimensional operators involving both representations and insertions of the adjoint $\Phi_{15}$ with inverse powers of $M_{\rm Pl}$ set lower bounds on the couplings in the scalar potential, and tend to favor a CP-breaking scale of $\vCP = {\cal O}(10^{11})$ GeV. Achieving consistency with the viable parameter space shown in Fig.~\ref{fig:plot} then requires, at least, a tuning at the percent level in the scalar potential. Radiatively generated mass terms in the spirit of the so-called finite naturalness~\cite{Farina:2013mla,FileviezPerez:2023rxn} intensify the tuning. 

In the scenario where a GUT scale is considered, the tension described above disappears but analogous tensions are expected between the scalars and radiative corrections from the bosonic fields with masses at the GUT scale.

A final comment concerns the implications of this hierarchy of scales for baryogenesis. Generating a matter-antimatter asymmetry is particularly non-trivial in theories where CP is a symmetry of the UV Lagrangian~\cite{Murgui:2025scx}. In the present framework, heavy right-handed neutrinos are present and a CP-violating phase is generated at the scale $\vCP = 10^9$ GeV. This opens the possibility of realizing leptogenesis, although its detailed implementation depends on the right-handed-neutrino spectrum and on the size of the CP-violating invariants in the lepton sector.

If a hierarchical spectrum is realized in the heavy neutrino sector, the Dirac Yukawa couplings, which are fixed by the up-quark sector accordingly to the $\SU(4)$ relation for the mass matrices $\mathsf{M}_\nu^D  = \mathsf{M}_u$, may still allow for leptogenesis via the out-of-equilibrium decay of the lightest heavy neutrino, provided its decay rate is suppressed by $m_u^2$. In the high-scale quark-lepton unification scenario, the Dirac neutrino mass matrices are real. Nevertheless, the CP-violating phase can be transmitted to the heavy Majorana neutrino sector through the mixing of $\phi_4^0$ and $\phi_{10}^0$ (see Appendix~\ref{app:potential} for details). 
Yet, the resulting CP asymmetry is expected to be suppressed, since the physical phase originates from the interplay between the renormalizable coupling $\mathsf{Y}_{\nu_R}$ and the effective Yukawa generated from higher dimensional operators,
\begin{equation}
 Y_{\nu_R} \Phi_{10}^* F_{u\nu}^T C F_{u\nu} 
+  \frac{1}{\Lambda} \Phi_{10}^* F_{u\nu}^T C \Phi_{15} F_{u\nu}  + \text{h.c.}~.
\end{equation}
Parametrically, this leads to a suppression of the CP asymmetry of order $y_d = {\cal O}(10^{-5})$ in the CP asymmetry. By contrast, in the GUT scenario the CP-violating phase enters in the neutrino mass matrix through the mixing between $\nu_R$ and the new fermionic gauge singlets $N$. In this case, an ${\cal O}(1)$ CP-violating invariant in the heavy neutrino sector can be expected. A dedicated study of leptogenesis, or of alternative for baryogenesis mechanisms within this framework, is left for future work. 

At present, a suppressed Dirac CP phase in the PMNS matrix is not decisively excluded. Although global fits and long-baseline appearance data tend to prefer values of $\delta_{\rm CP}$ near $-\pi/2$, the result remains correlated with the neutrino 
mass ordering, $\theta_{23}$, and experimental systematics~\cite{Esteban:2024eli,T2K:2023smv,NOvA:2021nfi}. In addition, T2K and NO$\nu$A show a mild tension in their preferred regions of 
$\delta_{\rm CP}$~\cite{T2K:2025wet}, and CP-conserving values for normal ordering remain allowed in parts of the parameter space~\cite{Esteban:2024eli,T2K:2025wet}. Future data from T2K,
NO$\nu$A, Hyper-Kamiokande, and DUNE may therefore help discriminate between scenarios with suppressed and unsuppressed low-energy leptonic CP violation.

\section{Concluding remarks}
\label{sec:remarks}

In this paper we have presented a UV completion of the minimal simplified model in which the strong CP problem is solved {\it \`a la} Nelson-Barr. In this theory, each generation of the Standard Model quarks and leptons is unified into representations of $\SU(4)$. This symmetry, which predicts the equality of the SM down quark and charged lepton masses at the quark-lepton unification scale, plays a crucial role in enforcing the texture required to kill contributions to $\bar \theta_{\rm QCD}$ from quark masses at tree-level. This is achieved once the fermionic sector is extended in the minimal non-trivial way, namely, by the inclusion of an antisymmetric representation, which contains a vector-like down-type quark.

The vector-like down quark plays a two-fold role: it enables a realistic  description of the second and third generation of fermion masses, while simultaneously providing the minimal ingredient required for the implementation of the Nelson-Barr mechanism. 

In contrast to minimal realizations of quark-lepton unification, baryon number is not conserved in this theory. The combined requirements of a high-quality Nelson-Barr and the non-observation of baryon number violation significantly restrict the viable parameter space, fixing the scale of spontaneous CP violation and the new vector-like down quark to be around $10^9$ GeV.

Although the scale of the new states is too high to produce observable effects in flavor observables, the theory predicts a distinctive and, in principle, experimentally accessible signal of baryon number violation: the decay mode $n\to K^+ \ell^-$, with $\ell = e,\mu$. The predicted rate lies within the reach of upcoming nucleon decay and neutrino experiments. Since the existing bound on this channel dates back to Fr\'ejus, it would be especially interesting to revisit this mode with Super-Kamiokande data, where a dedicated treatment of the relevant backgrounds could substantially improve the present sensitivity.

The relation between the quark-lepton unification scale and the scale suppressing the non-renormalizable operators responsible for down-quark masses leads to two possible scenarios in the UV. If these operators are suppressed by the Planck
scale, the observed down-quark masses point to a high quark-lepton unification scale of $\Lambda_{\rm QL}={\cal O}(10^{14})$ GeV. Alternatively, if the effective cutoff is identified with an intermediate GUT scale, the framework can be embedded in a grand-unified theory for $\Lambda_{\rm QL} = {\cal O}(10^{11})$ GeV. The two scenarios may also lead to different expectations for low-energy leptonic CP violation, which could be tested by future long-baseline experiments such as T2K, NO$\nu$A, Hyper-Kamiokande and DUNE.

Overall, this theory provides a concrete and instructive example of how the Nelson-Barr mechanism can arise naturally in well-motivated extensions of the Standard Model, while leading to sharp phenomenological predictions that may be tested in the near future.

\begin{acknowledgements}
First and foremost, I would like to express my deepest thanks to Mark B. Wise for his substantial contributions to this work. In particular, he pointed out that this theory predicts a single baryon number violating dominant decay mode, 
$n \to K^+ (e,\mu)^-$, and provided invaluable insights through numerous discussions on the blackboards of the fourth floor of Downs-Lauritsen lab, as well as over wine and cheese at the Ath. This work is dedicated to his memory.

I also thank Pavel Fileviez Perez for earlier collaboration on both quark-lepton unification and Nelson–Barr topics.

I am grateful to Javier Fuentes, Admir Greijo, Gino Isidori, Claudio Andrea Manzari, and Quoc Viet Nguyen for useful discussions, and especially to Hitoshi Murayama for conversations regarding DUNE. I thank LBNL and UC Berkeley for hospitality during the completion of this work, and the participants of the Model Building in 2030 workshop at the IAS for stimulating discussions. I also thank the referee for very useful and constructive comments.
\end{acknowledgements}

\appendix 

\section{A bit of algebra}
In this appendix we play around the diagonalization of ${\cal M}_d{\cal M}_d^\dagger$.
\subsection{Bounded eigenvalues}
\label{app:evaluebound}

The mass matrix in Eq.~\eqref{eq:M2}, $\mathsf{M}^2$, can be rewritten as\footnote{In this section we use Dirac notation to streamline the distinction between vectors, matrices, and c-numbers.}
\begin{equation}\label{eq:M2red2}
    \mathsf{M}^2 = \mathsf{D} - |L\rangle \langle L|~,
\end{equation}
where $\mathsf{D}= \kappa_M^2\, \text{diag}(m_e^2,m_\mu^2,m_\tau^2)$ and $|L\rangle= \kappa_M (m_e \epsilon_1, m_\mu \epsilon_2, m_\tau \epsilon_3)^T$, with $\epsilon_i \equiv \mu_i/M_D$. This form makes explicit that $\mathsf{M}^2$ differs from a diagonal matrix by a rank-one operator. 

The spectrum of $\mathsf{M}^2$ is the set of real numbers $\lambda$ satisfying $\det(\mathsf{M}^2-\lambda \mathbb{I}) = 0$. 
Using the matrix determinant lemma, and assuming $\lambda \neq \kappa_M^2  m_{\ell_i}^2$, one finds
\begin{equation}\label{eq:identity}
\det(\mathsf{M}^2 - \lambda \mathbb{I}) = \det(\mathsf{D}- \lambda \mathbb{I}) f(\lambda) ~,
\end{equation}
where
\begin{equation}
\begin{split}
    f(\lambda) &= ( 1 - \langle L | (\mathsf{D}-\lambda \mathbb{I})^{-1} |L \rangle)\\
               &= \bigg( 1 - \sum_{i=1}^3 \frac{|L_i|^2}{(\kappa_M^2 m_{\ell_i}^2 - \lambda)}\bigg)~.
\end{split}
\end{equation}
Therefore, Eq.~\eqref{eq:identity} tells us that, for $\lambda \neq \kappa_M^2 m_{\ell_i}^2$, the eigenvalues of $\mathsf{M}^2$ are given by the zeros of $f(\lambda)$. 

The function $f(\lambda)$ has the following properties:
\begin{itemize}
    \item[(i)] Asymptotically, $\lim_{\lambda \to \pm \infty} f(\lambda) = 1$.
    \item[(ii)] Near each pole $\lambda = \kappa_M^2 m_{\ell_i}^2$, $$\lim_{\lambda \to (\kappa_M^2 m_{\ell_i}^2)^\pm}f(\lambda) = \pm \infty.$$
    \item[(iii)] $f(\lambda)$ is strictly monotonically decreasing on each interval of its domain separated by the poles, 
   $$ f'(\lambda) = -\sum_{i=1}^3 \frac{|L_i|^2}{(\kappa_M^2 m_{\ell_i}^2 - \lambda)^2} <0~.$$
\end{itemize}
Hence, as the diagram below shows, 
\begin{equation*}
\begin{tikzpicture}[
  x=\columnwidth/9, y=1cm,
  >=Latex,
  tick/.style={line width=0.45pt},
  pole/.style={densely dashed, line width=0.45pt},
  root/.style={circle, draw, fill=white, inner sep=1.2pt, line width=0.45pt},
  lab/.style={font=\small},
  sign/.style={font=\small},
]
\def\dOne{1.6}
\def\dTwo{3.6}
\def\dThree{5.6}
\def\dminfty{0}
\def\dpinfty{7}
\def\mOne{0.9}
\def\mTwo{2.6}
\def\mThree{4.7}
\def\mFour{6.5}
\draw[->, line width=0.55pt] (0,0) -- (7.4,0) node[right, lab] {$\lambda$};
\foreach \x/\labtxt in {\dOne/$\kappa_M^2 m_e^2$, \dTwo/$\kappa_M^2 m_\mu^2 $, \dThree/$\kappa_M^2 m_\tau^2$}{
  \draw[pole] (\x,0) -- (\x,-1.35);
  \draw[tick] (\x,0.10) -- (\x,-0.10);
  \node[above, lab] at (\x,0.12) {\labtxt};
}
\foreach \x/\labtxt in { \dminfty/$-\infty$, \dpinfty/$+\infty$}{
  \node[above, lab] at (\x,0.12) {\labtxt};
}
\foreach \x/\labtxt in { \mOne/zero, \mTwo/zero, \mThree/zero, \mFour/$>0$}{
  \node[above, lab] at (\x,-1.6) {\labtxt};
}
\node[sign] at (\dminfty+0.4,-0.40) {$+$};
\node[sign] at (\dOne-0.4,-0.40) {$-$};
\node[sign] at (\dOne+0.4,-0.40) {$+$};
\node[sign] at (\dTwo-0.4,-0.40) {$-$};
\node[sign] at (\dTwo+0.4,-0.40) {$+$};
\node[sign] at (\dThree-0.4,-0.40) {$-$};
\node[sign] at (\dThree+0.4,-0.40) {$+$};
\node[sign] at (\dpinfty,-0.40) {$+$};
\draw[->, line width=0.45pt] (\dminfty+0.4,-0.8) -- (\dOne-0.4,-1.1);
\draw[->, line width=0.45pt] (\dOne+0.6,-0.8) -- (\dTwo-0.6,-1.1);
\draw[->, line width=0.45pt] (\dTwo+0.6,-0.8) -- (\dThree-0.6,-1.1);
\draw[->, line width=0.45pt] (\dThree+0.6,-0.8) -- (\dpinfty-0.1,-1.1);
\node[left, lab] at (0,-0.75) {$f(\lambda)$};
\end{tikzpicture}
\end{equation*}
properties (i), (ii) and (iii) imply that $f(\lambda)$ has exactly one zero in each interval between successive poles. Consequently, the eigenvalues of $\mathsf{M}^2$ interlace with those of $\mathsf{D}$, yielding 
\begin{equation}\label{eq:eigenlemma}
    m_d^2 \leq \kappa_M^2 m_e^2 \leq m_s^2 \leq \kappa_M^2 m_\mu^2 \leq m_b^2 \leq  \kappa_M^2 m_\tau^2~.
\end{equation}
Equalities occur precisely when $\kappa_M^2 m_{\ell_i}^2$ is itself an eigenvalue of $\mathsf{M}^2$. In particular, this happens if and only if $L_i = 0$ for non-degenerated $m_{\ell_i}^2$.

\subsection{Texture of the diagonalizing matrix}
\label{app:diagonalizer}
The unitary matrix that diagonalizes $\mathsf{M}^2$ is defined by 
\begin{equation}
\mathsf{U}_L^\dagger \mathsf{M}^2 \mathsf{U}_L = \text{diag}(\lambda^{(1)}, \lambda^{(2)}, \lambda^{(3)})~.
\end{equation}
The eigenvalues $\lambda^{(k)}$, as discussed in the main text, correspond to the SM down-type quark masses squared. In this section, we derive the texture of $\mathsf{U}_L$ by exploiting the strong hierarchy $m_e \ll m_\mu \ll m_\tau$.

We begin by rewriting $\mathsf{M}^2$ in the form given in Eq.~\eqref{eq:M2red2}.
From the eigenvalue equation, $\mathsf{M}^2 |\mathsf{m}^{(k)}\rangle = \lambda^{(k)} |\mathsf{m}^{(k)}\rangle $, we obtain the component-wise relation
\begin{equation}
(\kappa_M^2 m_{\ell_i}^2 -\lambda^{(k)}) \mathsf{m}^{(k)}_i =   \langle L | \mathsf{m}^{(k)}\rangle L_i ~.
\end{equation}
Here $|\mathsf{m}^{(k)}\rangle$ denotes the normalized eigenvector of $\mathsf{M}^2$ associated with the eigenvalue $\lambda^{(k)}$. It therefore corresponds to the $k$-th column of the diagonalizing matrix $\mathsf{U}_L$. 

We now assume that $\langle L | \mathsf{m}^{(k)}\rangle \neq 0$.\footnote{As discussed in the previous section, if $\langle  L | \mathsf{m}^{(k)} \rangle = 0$, then the corresponding eigenstate is completely decoupled from the rank-one perturbation and the eigenvalue remains fixed at $\kappa_M^2 m_{\ell_i}^2$ with $L_i=0$.} In this case, the eigenvector components satisfy
\begin{equation}\label{eq:alignment}
\mathsf{m}^{(k)}_i \propto \frac{\kappa_M m_{\ell_i} \epsilon_i }{\kappa_M^2 m_{\ell_i}^2 - \lambda^{(k)}}~,
\end{equation}
where the proportionality constant is given by $\langle L | \mathsf{m}^{(k)}\rangle$. Equation~\eqref{eq:alignment} strongly constrains the structure of the diagonalizing matrix $\mathsf{U}_L$: each column is aligned with the vector $|L\rangle$, with relative weights determined by the denominators $\kappa_M^2 m_{\ell_i}^2 - \lambda^{(k)}$. As a consequence,  
the ratios $\mathsf{m}^{(k)}_i / \mathsf{m}^{(k)}_j$ are fully fixed by the charged lepton masses, down type quark masses, and the parameters $\epsilon_i$. 

Let us examine more closely the structure of $\mathsf{U}_L$ by exploiting our knowledge of the eigenvalues of $\mathsf{M}^2$. Focusing on the third column, Eq.~\eqref{eq:alignment} implies that
\begin{equation}
\frac{\mathsf{m}^{(3)}_{i=1,2}}{\mathsf{m}^{(3)}_3}\simeq \frac{m_{\ell_i} \epsilon_i}{m_\tau \epsilon_3} \frac{\kappa_M^2 m_\tau^2 - \lambda^{(3)}}{\kappa_M^2 m_{\ell_i}^2 - \lambda^{(3)}}~.
\end{equation}
The largest eigenvalue satisfies $\lambda^{(3)} \simeq \kappa_M^2 m_\tau^2$. This is encouraged by the approximate bottom-tau Yukawa unification, since the ratio between the bottom and tau masses, $m_b(2\text{ GeV}) \simeq 2.35 \, m_\tau(2\text{ GeV})$ becomes close to unity when the Yukawa couplings are evolved to the quark-lepton unification scale. Therefore, $\lambda^{(3)}$ must differ from $\kappa_M^2 m_\tau^2$ by an order-one factor, so that $|\epsilon_3| ={\cal O}(1)$. Perturbation theory on $\mathsf{M}^2$ tells us that 
\begin{equation}
   \kappa_M^2 m_\tau^2 - \lambda^{(3)} = \kappa_M^2 m_\tau^2 |\epsilon_3|^2 + {\cal O}(\kappa_M^2 m_\mu^2)  ~,
\end{equation}
so that, using the hierarchy $m_e \ll m_\mu \ll m_\tau$,
\begin{equation}
\frac{\mathsf{m}^{(3)}_{i=1,2}}{\mathsf{m}^{(3)}_3}={\cal O}(1) \frac{m_{\ell_i}}{m_\tau}~.
\end{equation}

Proceeding similarly for the second column of $\mathsf{U}_L$, taking into account that $\lambda^{(2)} = m_s^2 = {\cal O}(\kappa_M^2 m_\mu^2)$ and using Eq.~\eqref{eq:alignment}, we find
\begin{equation}
\begin{split}
\frac{\mathsf{m}^{(2)}_1}{\mathsf{m}^{(2)}_2} &\simeq \frac{m_e \epsilon_1}{m_\mu \epsilon_2} \frac{ \kappa_M^2 m_\mu^2 - \lambda^{(2)}}{\kappa_M^2 m_e^2 - \lambda^{(2)}} = {\cal O}(1) \frac{m_e}{m_\mu}~,\\
\frac{\mathsf{m}^{(2)}_3}{\mathsf{m}^{(2)}_2} &\simeq \frac{m_\tau \epsilon_3}{m_\mu \epsilon_2} \frac{\kappa_M^2 m_\mu^2 - \lambda^{(2)}}{\kappa_M^2 m_\tau^2-\lambda^{(2)}} = {\cal O}(1) \frac{m_\mu}{m_\tau}~.
\end{split}
\end{equation}

Finally, for the first column, although $\lambda^{(1)}$ cannot be directly identified with the down-quark mass (as discussed in the main text), we can still use the bound $\lambda^{(1)} \leq \kappa_M^2 m_e^2$ derived in the previous section. It then follows that 
\begin{equation}
\begin{split}
\frac{\mathsf{m}^{(1)}_{i=2,3}}{\mathsf{m}^{(1)}_1} &\simeq \frac{m_{\ell_i} \epsilon_i}{m_e \epsilon_1} \frac{m_e^2 - \lambda^{(1)}}{m_{\ell_i}^2 - \lambda^{(1)}} = {\cal O}(1) \frac{m_e}{m_{\ell_i}}~.
\end{split}
\end{equation}
We therefore conclude that the diagonalizing matrix $\mathsf{U}_L$ is approximately diagonal, with off-diagonal entries parametrically suppressed by ratios of the charged lepton masses:
\begin{equation}\label{eq:textureUL}
\mathsf{U}_L = \begin{pmatrix} {\cal O}(1) & {\cal O}\left( \frac{m_e}{m_\mu}\right) & {\cal O}\left( \frac{m_e}{m_\tau} \right) \\
{\cal O}\left(\frac{m_e}{m_\mu}\right) & {\cal O}(1)& {\cal O}\left( \frac{m_\mu}{m_\tau}\right) \\
{\cal O}\left( \frac{m_e}{m_\tau}\right) & {\cal O}\left( \frac{m_\mu}{m_\tau}\right) & {\cal O}(1)\end{pmatrix}~.
\end{equation}

For completeness, we briefly discuss the fate of the CKM phase. At first sight, one might worry that the structure of Eq.~\eqref{eq:M2red2} could eliminate all CP-violating phases. Indeed, under a redefinition of the left-handed down quark fields, 
\begin{equation}
    d_{L} \to \mathsf{P} \, d_L \quad \text{with }\ \mathsf{P} = \text{diag}(\e^{\iu \alpha_1}, \e^{\iu \alpha_2}, \e^{\iu \alpha_3})~,
\end{equation}
the mass matrix transforms as 
\begin{equation}
\mathsf{M}^2 \to \mathsf{P}^\dagger \mathsf{M}^2 \mathsf{P} = \mathsf{D}^2 - (\mathsf{P}^\dagger |L\rangle )(\mathsf{P}^\dagger |L\rangle)^\dagger~,
\end{equation}
where we have used that $\mathsf{P}$ and $\mathsf{D}^2$ are diagonal matrices and therefore commute. As a result, the three phases $\alpha_i$ can be chosen to absorb the phases of the parameters $\epsilon_i = |\epsilon_i|\e^{-\iu \alpha_i}$, rendering $\mathsf{M}^2$ real. 
In other words, one can always choose a weak basis in which the light down-quark sector of $\mathsf{M}^2$ is CP conserving, and the diagonalizing matrix $\mathsf{U}_L$ is real and orthogonal. 

However, the transformation $\mathsf{P}$ acts on the entire $Q_L$ doublet. In the same basis, the up-quark mass matrix transforms as
\begin{equation}
\mathsf{M}_u \to \mathsf{P}^\dagger \mathsf{M}_u~.
\end{equation}
Therefore, even if $\mathsf{M}_u$ is initially real, it becomes complex after redefinition. The phases then cannot be removed simultaneously from both quark sectors and, therefore, a physical CKM phase remains.

We can now check whether the texture of $\mathsf{U}_L$ in Eq.~\eqref{eq:textureUL} is sufficient to reproduce the observed magnitude of the CKM matrix elements. Comparing  Eq.~\eqref{eq:VCKM} with the structure of $\mathsf{U}_L$, one can infer the parametric size of the entries of the orthogonal matrix $\mathsf{O}_{uL}$. This leads to the schematic texture
\begin{equation}
    \mathsf{O}_{uL} = \begin{pmatrix} \scalebox{1}{$\blacksquare$} & \scalebox{0.7}{$\blacksquare$} & \centerdot \\ \scalebox{0.7}{$\blacksquare$}& \scalebox{1}{$\blacksquare$} & \scalebox{0.7}{$\blacksquare$} \\ \centerdot & \scalebox{0.7}{$\blacksquare$} & \scalebox{1}{$\blacksquare$} \end{pmatrix}~,
\end{equation}
where 
$$\centerdot  = {\cal O}(0.01), \quad \scalebox{0.7}{$\blacksquare$} = {\cal O}(0.1), \quad \text{and} \quad \blacksquare= {\cal O}(1)~.$$ ~
Decomposing the CKM matrix as 
\begin{equation}
\mathsf{V}_{\rm CKM} = \overline{ \mathsf{V}}_{\rm CKM} + \delta \mathsf{V}_{\rm CKM},
\end{equation}
where $\overline{\mathsf{V}}_{\rm CKM} = \mathsf{O}_{uL}^T \text{diag}(\mathsf{U}_L) \mathsf{O}_L$, and $\delta \mathsf{V}_{\rm CKM}$ contains the off-diagonal entries of $\mathsf{U}_L$ suppressed by the ratios of charged lepton masses multiplied by entries of $\mathsf{O}_{uL}$ and $\mathsf{O}_L$. 
Although $\overline{\mathsf{V}}_{\rm CKM}$ may contain complex phases, these correspond to row- and column rephasings and therefore do not contribute to the Jarlskog invariant, $J$. Consequently, the leading contributions to $J$ come from interference terms involving at least one insertion of $\delta \mathsf{V}_{\rm CKM}$. Since these contain the suppressed off-diagonal elements of $\mathsf{U}_L$, order-one phases in $\mathsf{U}_L$ are required to reproduce its experimental value of $J \simeq 3 \times 10^{-5}$~\cite{ParticleDataGroup:2024cfk}.

\section{Scalar mixing}
\label{app:potential}
There is a choice of basis in the scalar potential where the vacuum expectation value of one of the $\Phi_{4\, a,b}$ scalar fields is real, so that the physical CP-violating phase $\e^{\iu \theta_{4}}$ resides entirely in a single $\Phi_4$ field. Let us consider, for simplicity, the scalar $\Phi_4$ with a real vev.

The most general renormalizable  scalar potential involving the electromagnetically neutral components of $\Phi_4$ and $\Phi_{10}$, consistent with the UV symmetries of the theory, ${\cal G}_{\rm PS}$, can then be written as 
\begin{equation}
\begin{split}
\!\! V &= - \mu_4^2 |\phi_4^0|^2 - \mu_{10}^2 |\phi_{10}^0|^2 + \tfrac{\lambda_4}{2} |\phi_4^0|^4 + \tfrac{\lambda_{10}}{2}|\phi_{10}^0|^4 \\
& \ + \tfrac{\lambda_{4 \, 10}}{2} |\phi_4^0|^2 |\phi_{10}^0|^2 + \tfrac{f}{2\sqrt{2}}\big( \phi_{10}^{0*}(\phi_4^0)^2    + \text{h.c} \big).
\end{split}
\end{equation}

Assuming the standard normalization $\langle \phi_4^0 \rangle = \vCP / \sqrt{2}$ and $\langle \phi_{10}^0 \rangle = v_{10} /\sqrt{2}$ as defined in the main text, the minimization conditions of the potential are
\begin{eqnarray}
-4 \mu_{10}^2 v_{10} + 2 \lambda_{10}v_{10}^3 + (f + \lambda_{4\, 10} v_{10}) \vCP^2 =  0~, \label{eq:conditionI}\\
-4\mu_4^2 +2f v_{10} + \lambda_{4\, 10} v_{10}^2  + 2 \lambda_4 \vCP^2 = 0~.\label{eq:conditionII}
\end{eqnarray}
Given that $v_{10} \gg \vCP$, condition \ref{eq:conditionI} fixes $v_{10} = \sqrt{2/\lambda_{10}} \, \mu_{10}$ in a natural way. Condition~\ref{eq:conditionII}, however, implies that the parameters $f$ and $\lambda_{4\, 10}$ must be small to establish a hierarchical lower CP-breaking scale. 

In the absence of additional symmetries, these parameters are generically bounded from below by Planck-suppressed operators induced by gravity, such as 
\begin{equation}
\begin{split}
&\frac{1}{M_{\rm Pl}} \Phi_4^\dagger \Phi_{15} \Phi_4 \Phi_{10}^\dagger \Phi_{10} + \frac{1}{M_{\rm Pl}} \Phi_{10}^\dagger \Phi_{15}^2 \Phi_4^2 \\
&\quad + \frac{1}{M_{\rm Pl}}\Phi_4^\dagger \Phi_4 \text{Tr}\{\Phi_{15}^3\} +\cdots ~.
\end{split}
\end{equation}
Parametrically these contributions induce 
\begin{equation}
    f \sim \frac{\Lambda_{\rm LQ}^2}{M_{\rm Pl}}, \quad \lambda_{4\, 10}  = \frac{\Lambda_{\rm LQ}}{M_{\rm Pl}}, \quad \text{and} \quad \mu_4^2 = \frac{\Lambda_{\rm LQ}^3}{M_{\rm Pl}}~.
\end{equation} 
Given that $v_{10} \sim \Lambda_{\rm LQ}$, substituting these estimates into Eq.~\eqref{eq:conditionII} yields a CP-breaking scale $\vCP \sim {\cal O}(\Lambda_{\rm LQ} \sqrt{\Lambda_{\rm LQ}/M_{\rm Pl}}) \sim 10^{11} \text{ GeV}$, which lies approximately one order of magnitude above what the maximum value allowed by the quality of the Nelson-Barr mechanism (see Eq.~\eqref{eq:qualityB}).\footnote{A similar result is obtained if $v_{10}$ and $v_{\rm CP}$ are determined using Eqs.~\eqref{eq:conditionII} and~\eqref{eq:conditionI}, respectively, the latter with parameters generated by higher dimensional operators.} Reconciling these scales therefore requires a tuning at the percent level in the scalar potential.

Turning to the masses of the neutral scalars in $\Phi_4$ and $\Phi_{10}$, after imposing the minimization conditions, the mass matrix for the CP-even scalar fields takes the form
\begin{equation}
v_{10}^2
\begin{pmatrix} 
 \lambda_4 \, \theta_\Phi^2 & \tfrac{1}{2} (\frac{f}{v_{10}} + \lambda_{4\, 10}) \, \theta_\Phi \\
\tfrac{1}{2}(\tfrac{f}{v_{10}} + \lambda_{4\, 10})  \, \theta_\Phi & \lambda_{10}  - \tfrac{1}{4} \frac{f}{v_{10}} \theta_\Phi^2 
\end{pmatrix}~,
\end{equation}
where $\theta_{\Phi} \equiv \vCP/v_{10}$. Using the gravity-induced estimates above, one finds  $f/v_{10} \sim \lambda_{4\, 10} \sim \Lambda_{\rm LQ}/M_{\rm Pl} \sim  y_d$, with $y_d$ the down-quark Yukawa coupling. The two small expansion parameters of the system are therefore $\theta_\Phi \sim y_d \sim 10^{-5}$.

At leading order in these parameters, perturbation theory gives
\begin{equation}\label{eq:massesV}
M_{\Phi_4}^2 \sim  \lambda_4  v_{\rm CP}^2, \quad  \text{and} \quad M_{\Phi_{10}}^2 \sim \lambda_{10} v_{10}^2~,
\end{equation}
with $\theta_\Phi$ quantifying the mixing angle between the two scalar fields. An analogous mass spectrum and mixing angle are expected for the scalar leptoquarks $\phi_4^\alpha$ and $\phi_{10}^\alpha$. 

We note that the tension in the scalar potential to achieve the gap between $\Lambda_{\rm QL}$ and $\vCP$ worsens once radiative corrections to $M_{\Phi_4}^2$ are taken into account. In particular, heavy gauge bosons such as the vector leptoquarks $X^\mu$ and the neutral $Z'$ associated with the breaking of $\SU(4)_C$ and $\mathrm{U}_{B-L} \otimes \mathrm{U}_R$, respectively, generate finite contributions of the form~\cite{FileviezPerez:2023rxn}
\begin{equation}
    \delta M_{\Phi_4}^2|_{{\rm 1-loop}} \sim \frac{\alpha_s}{4\pi} \Lambda_{\rm LQ}^2 \ln \left(\frac{\Lambda_{\rm LQ}^2}{M_{\Phi_4}^2}\right)~,
\end{equation}
which naturally suggest $M_{\Phi_4}={\cal O}(10^{12} \text{ GeV})$. 

\section{Running of the gauge couplings}
\label{app:RGEs}
At one-loop, the running of the gauge couplings is given by
\begin{equation}
    \alpha_i^{-1}(\mu) = \alpha_i^{-1}(M_{\rm ref}) - \frac{b_i}{2\pi} \ln \left( \frac{\mu}{M_{\rm ref}}\right) ~,
\end{equation}
where $b_i$ denotes the one-loop beta-function coefficient appropriate to the particle content between the scale $M_{\rm ref}$ and the scale $\mu$.

The symmetry-breaking chain relevant for the running is
\begin{equation}
\begin{split}
&\SU(4)_C\times\SU(2)_L\times\U(1)_R\\
&\ \xrightarrow{\langle\Phi_{15}\rangle}
\SU(3)_C\times\SU(2)_L\times\U(1)_R\times\U(1)_{B-L} \\
&\ \xrightarrow{\langle\Phi_{10}\rangle\ {\rm or}\ \langle\Phi_{4}\rangle}\SU(3)_C\times\SU(2)_L\times\U(1)_Y .
\end{split}
\end{equation}
Accordingly, above the scale $\langle\Phi_{15}\rangle \equiv \Lambda_{\rm QL}$, the couplings to be run are $\alpha_4$, $\alpha_2$ and
$\alpha_R$. Between $\Lambda_{\rm QL}$ and the scale at which the $\U(1)_R\times\U(1)_{B-L}$ symmetry is broken to hypercharge, denoted by
$\Lambda_{\rm SM}$, the theory is described by the $3211$ gauge group, 
\begin{equation}
{\cal G}_{3211} = \SU(3)_C \otimes \SU(2)_L \otimes \U(1)_R\otimes \U(1)_{B-L} ,
\end{equation}
and the
couplings are $\alpha_3$, $\alpha_2$, $\alpha_R$ and $\alpha_{B-L}$. Below
$\Lambda_{\rm SM}$, the running is that of the SM gauge couplings
$\alpha_3$, $\alpha_2$ and $\alpha_Y$, up to the additional thresholds specified
below.

At the first threshold, $\Lambda_{\rm QL}$, the matching conditions are
\begin{equation}
\begin{split}
\alpha_3(\Lambda_{\rm QL})&=\alpha_4(\Lambda_{\rm QL}),
    \\
    \alpha_{B-L}(\Lambda_{\rm QL})
    &=
    \frac{3}{2}\alpha_4(\Lambda_{\rm QL}) .
\end{split}
\end{equation}
Here the normalization is chosen according to Eq.~\eqref{eq:Ynorm}.
At the second threshold, $\Lambda_{\rm SM}$, the hypercharge coupling is matched
according to
\begin{equation}
    \alpha_Y^{-1}(\Lambda_{\rm SM})
    =
    \alpha_R^{-1}(\Lambda_{\rm SM})
    +
    \alpha_{B-L}^{-1}(\Lambda_{\rm SM})\, .
\end{equation}
In the limit in which the two thresholds are taken to be degenerate,
$\Lambda_{\rm QL}=\Lambda_{\rm SM}$, the intermediate $3211$
regime has no logarithmic running. In that case the matching reduces to the
combined condition
\begin{equation}
    \alpha_Y^{-1}(\Lambda_{\rm QL})
    =
    \alpha_R^{-1}(\Lambda_{\rm QL})
    +
    \frac{2}{3}\alpha_4^{-1}(\Lambda_{\rm QL}).
\end{equation}
The $\SU(2)_L$ factor is unaffected by these symmetry-breaking steps, and therefore $\alpha_2$ is continuous across the thresholds.

The $b_i$ coefficients at 1-loop are fixed by group theory and given by
\begin{equation}
\begin{split}
    b_i &= -\frac{11}{3} C_2(G_i) + \frac{2}{3} \sum_f T_i(R_f) \prod_{j\neq i} d_j(R_f) \\
    &\quad + \frac{1}{3} \sum_s T_i(R_s) \prod_{j\neq i} d_j(R_s),
\end{split}
\end{equation}
where $C_2(G_i)$ is the quadratic Casimir of the adjoint representation, $T_i(R)$ is the Dynkin index of the representation $R$ under the group factor ${\cal G}_i$, and $d_{j\neq i}$ denotes the multiplicity of the representation under the remaining group factors. The sums run over Weyl fermions ($f$) and complex scalars ($s$).

The beta-function coefficients depend on the energy interval. Below $v_{\rm CP}$
the running is that of the SM, $b_Y^{\rm SM}=41/6$, $b_2^{\rm SM}=-19/6$, $b_3^{\rm SM}=-7$.

For the high-scale quark-lepton unification scenario, between $v_{\rm CP}$ and $\Lambda_{\rm SM} \sim \Lambda_{\rm QL}$,
the vector-like down quarks in $F_6$ and the two copies of $\Phi_4$ contribute to the SM beta
functions. We denote the corresponding coefficients by
\begin{equation}
\begin{split}
b_Y^{\rm CP} &=
b_Y^{\rm SM}
+ \Delta b_Y\big|_{F_6}
+ 2\,\Delta b_Y\big|_{\Phi_4}\, ,\\
b_2^{\rm CP} &= b_2^{\rm SM}\, ,\\
b_3^{\rm CP} &=
b_3^{\rm SM}
+ \Delta b_3\big|_{F_6}
+ 2\,\Delta b_3\big|_{\Phi_4}\, .
\end{split}
\end{equation}
For the field content considered here, $F_6 = D_L \oplus (D^c)_L$, and $\Phi_4 = \Phi_4^\alpha \oplus \Phi_4^0$, so that
$\Delta b_Y\big|_{F_6} =\Delta b_Y\big|_{\Phi_4} =4/9$, $\Delta b_3\big|_{F_6}= 2/3$, $\Delta b_3\big|_{\Phi_4}=1/6$.
Above $\Lambda_{\rm QL}$, the running is performed in the ${\cal G}_{\rm QL}$ theory, with coefficients
\begin{equation}\label{eq:b4coeff}
    b_4^{\rm QL} = -8, \ b_2^{\rm QL} = -\frac{19}{6}, \ b_R^{\rm QL} =\frac{49}{6} .
\end{equation}

In the low-scale quark-lepton scenario, there is an intermediate regime with
gauge group ${\cal G}_{3211}$. In this interval the relevant coefficients are
\begin{equation}
\begin{split}
   b_3^{3211} &= -6,\ b_2^{3211} = -\frac{19}{6}, \\
  b_R^{3211} &= \frac{49}{6},\ b_{B-L}^{3211} = \frac{10}{3} .
\end{split}
\end{equation}
The coupling $\alpha_{B-L}$ is therefore run independently in this scenario. The
hypercharge coupling is recovered only at the lower threshold through the
matching condition above.

Above $\Lambda_{QL}$, the coefficients in Eq.~\eqref{eq:b4coeff} change by subtracting the contribution from $\Phi_{10} \sim (10,1,1)$, i.e. $\Delta b_4^{\rm QL}\big|_{\Phi_{10}} = 1$, $\Delta b_2^{\rm QL} \big|_{\Phi_{10}} = 0$, and $\Delta b_R^{\rm QL} \big|_{\Phi_{10}} = 10/3$.

\vfill

\bibliography{ref.bib}
\end{document}